\documentclass[aps,prd,preprint,a4paper,showpacs,nofootinbib,superscriptaddress]{revtex4-2}
\usepackage{bm}
\usepackage{indentfirst}
\usepackage{amsmath}
\usepackage{graphicx}
\usepackage{float}
\usepackage{amssymb}
\usepackage{subfigure}
\usepackage{amssymb}
\usepackage{hyperref}
\usepackage{epstopdf}
\usepackage{graphicx}
\usepackage[section]{placeins}

\usepackage[utf8]{inputenc}
\hypersetup{
    colorlinks=true,
    linkcolor=red,
    citecolor=blue,
}
\usepackage{color}
\usepackage[T1]{fontenc}
\usepackage{txfonts}
\usepackage{orcidlink}

\usepackage{adjustbox,lipsum}

\newcommand{\Rmnum}[1]{\expandafter\@slowromancap\romannumeral #1@} 
\newcommand{\bq}{\begin{equation}}
\newcommand{\eq}{\end{equation}}
\newcommand{\bqn}{\begin{eqnarray}}
\newcommand{\eqn}{\end{eqnarray}}
\newcommand{\nb}{\nonumber}
\newcommand{\lb}{\label}

\makeatother

\begin{document}

\title{Continued fraction method for high overtone quasinormal modes in effective potentials with discontinuity}

\author{Guan-Ru Li}
\email[E-mail: ]{guanru.li@unesp.br}
\affiliation{Faculdade de Engenharia de Guaratinguet\'a, Universidade Estadual Paulista, 12516-410, Guaratinguet\'a, SP, Brazil}

\author{Jodin C. Morey}
\email[E-mail: ]{moreyj@lemoyne.edu}
\affiliation{Le Moyne College, Syracuse, New York, 13214-1301, USA}

\author{Wei-Liang Qian}
\email[E-mail: ]{wlqian@usp.br (corresponding author)}
\affiliation{Escola de Engenharia de Lorena, Universidade de S\~ao Paulo, 12602-810, Lorena, SP, Brazil}
\affiliation{Faculdade de Engenharia de Guaratinguet\'a, Universidade Estadual Paulista, 12516-410, Guaratinguet\'a, SP, Brazil}
\affiliation{Center for Gravitation and Cosmology, School of Physical Science and Technology, Yangzhou University, Yangzhou 225009, China}

\author{Ramin G. Daghigh}
\email[E-mail: ]{ramin.daghigh@metrostate.edu}
\affiliation{Natural Sciences Department, Metropolitan State University, Saint Paul, Minnesota, 55106, USA}

\author{Michael D. Green}
\email[E-mail: ]{michael.green@metrostate.edu}
\affiliation{Mathematics and Statistics Department, Metropolitan State University, Saint Paul, Minnesota, 55106, USA}

\author{Kai Lin}
\affiliation{Universidade Federal de Campina Grande, Campina Grande, PB, Brazil}

\author{Rui-Hong Yue}
\affiliation{Center for Gravitation and Cosmology, College of Physical Science and Technology, Yangzhou University, Yangzhou 225009, China}

\begin{abstract}
In this study, we extend Leaver's continued fraction method to evaluate black hole quasinormal modes (QNMs) in systems where the effective potential exhibits a discontinuity.  
Besides the low-lying modes, we particularly focus on high overtones, which are physically pertinent due to the substantial deformation of the QNM spectrum triggered by spectral instability.  
In our algorithm, we expand the wavefunction at the point of discontinuity, instead of the black hole horizon, and incorporate the Israel-Lanczos-Sen junction conditions.  
We apply this algorithm to compute the QNMs of the modified Regge-Wheeler potential up to $2000$ modes with high precision.  
For the low-lying modes, the numerical results show excellent agreement with those obtained using the matrix and Prony methods. 
The high overtones are significantly deformed, owing to the presence of echoes due to the discontinuity. 
This deformation in the asymptotic QNM spectrum reveals universal features that are largely independent of the specific form of the discontinuity in the potential, seemingly coinciding with those observed in the modified P\"oschl-Teller effective potential.  
We speculate on whether the  collective effect of the high overtones has an observational impact on gravitational wave signals.  

\end{abstract}

\date{Dec. 18th, 2024}

\maketitle


\newpage
\section{Introduction}\label{sec1}

Black holes stand as enigmatic pillars in the realm of theoretical physics, embodying gravity's most extreme manifestations.
The groundbreaking detection of gravitational waves from binary mergers by LIGO and Virgo~\cite{agr-LIGO-01, agr-LIGO-02, agr-LIGO-03, agr-LIGO-04} heralded a revolutionary era in observational astrophysics.
This achievement has catalyzed ambitious ongoing spaceborne programs such as LISA~\cite{agr-LISA-01}, TianQin~\cite{agr-TianQin-01}, and Taiji~\cite{agr-Taiji-01}, nurturing optimism for detailed  analysis of ringdown waveforms~\cite{agr-LISA-19, agr-LISA-20, agr-TianQin-05}.
At the heart of these ringdown waveforms lie the quasinormal modes (QNMs)~\cite{agr-qnm-review-02, agr-qnm-review-03, agr-qnm-review-06}, which have long served as a focal point of intensive research.
QNMs, which are independent of the perturbation source\cite{agr-qnm-30}, are crucial in gravitational wave astronomy, enabling researchers to pin down the properties of a black hole (mass, spin, and charge)~\cite{agr-BH-spectroscopy-review-04}.
 For that reason, QNMs are often referred to as a black hole's {\it fingerprint}.
Leaver's seminal work~\cite{agr-qnm-21, agr-qnm-29} established that QNMs correspond to poles in the Green's function of the master wave equation, while the branch cut governs late-time behavior, typically characterized by inverse power-law decay~\cite{agr-qnm-tail-01}.

Long-lived QNMs with low damping are expected to be observable in the gravitational waves emitted during the ringdown phase following black hole formation.  
Modes with very high damping, typically corresponding to high overtones, are unlikely to be detected directly, yet they have attracted considerable attention in recent years.  
A key reason is Hod's conjecture~\cite{agr-qnm-22}, which relates the behavior of the real part of the complex quasinormal frequencies with infinite damping rate to the quantization of black hole area and entropy through Bohr's correspondence principle.  
This proposal was initially supported by numerical results for the Schwarzschild spectrum obtained by Nollert~\cite{agr-qnm-continued-fraction-12} and by various analytic studies~\cite{agr-qnm-continued-fraction-23, agr-qnm-40}, but was later challenged by more detailed investigations.  
Maggiore~\cite{agr-qnm-23} subsequently introduced a refined formulation in which the modulus of the complex frequency, rather than the real part alone, determines the level spacing.
Among other developments, Babb {\it et al.}~\cite{agr-qnm-54} showed that, for spacetimes characterized by more than one intrinsic length scale, such as quantum-corrected black holes, the infinitely damped modes can in principle probe the geometry outside the horizon down to the smallest relevant scales, e.g., the quantum length scale.  
For a broader discussion of the theoretical significance of high overtones, we refer the reader to the review by Berti {\it et al.}~\cite{agr-qnm-review-03} and the references therein. 

The collection of all QNMs, high and low overtones, is referred to as the QNM spectrum.  
Recent studies have raised questions about the stability of QNM spectra under certain conditions, highlighting the ongoing complexity in black hole physics research~\cite{spectral-instability-review-20}.
Pioneered by Nollert and Price~\cite{agr-qnm-instability-02, agr-qnm-instability-03}, alongside Aguirregabiria and Vishveshwara~\cite{agr-qnm-27, agr-qnm-30}, it was revealed that even minute perturbations can dramatically alter higher-order QNM overtones.
This discovery challenged the prevailing notion that approximating the effective potential would yield only minor deviations in QNMs.
Subsequent investigations~\cite{agr-qnm-instability-11, agr-qnm-lq-03, agr-qnm-echoes-20} have suggested that even a minor discontinuity might induce a significant modification to the QNM spectrum's asymptotic behavior.
The resulting high-overtone modes are found to shift closer to the real frequency axis, contrasting with the typical ascent along the imaginary axis observed in most black hole metrics~\cite{agr-qnm-continued-fraction-12, agr-qnm-continued-fraction-23}.
For example, given two disjoint potential barriers, one of which is small, located at  $x_\mathrm{c}$, and playing the role of a perturbation introduced in an effective potential, the following asymptotic form of the $n$th quasinormal mode, $\omega_n$, has been obtained~\cite{agr-qnm-lq-03, agr-qnm-instability-65}
\bqn
\omega_n \sim \left(n+\frac12\right)\frac{\pi}{x_\mathrm{c}} - i\frac{\mathcal{C}}{x_\mathrm{c}}\ln\left[\left(n+\frac12\right)\frac{\pi}{x_\mathrm{c}}\right] + O(1) ,\label{asTruWKB}
\eqn
given the assumption that $\Re\omega_n\gg \Im\omega_n$.  
The constant $\mathcal{C}$ is of order unity, taking the value $1$ when the effective potential is composed of two delta functions, and $2$ when it consists of two rectangular barriers.
However, to the best of our knowledge, no general analytic result has been established when the magnitude of $\Im \omega_n$ cannot be neglected.
Specifically, for the P\"oschl-Teller potential truncated at $x_\mathrm{cut}$, one instead obtains an asymptotically ($n \gg 1$) linear relation between the real and imaginary parts of QNMs~\cite{agr-qnm-instability-65}
\bqn
\omega_n \sim \left(n+\frac12\right)\frac{\pi}{x_\mathrm{c}} - i\frac{1}{x_\mathrm{c}}\left\{\left(n+\frac12\right)\frac{\pi^2}{2\kappa x_\mathrm{c}}+\ln\left[\left(n+\frac12\right)\frac{\pi}{x_\mathrm{c}} \right]\right\} + O(1), \label{asTruWKB2}
\eqn
where $\kappa$ governs the exponential suppression in tortoise coordinate $r_*$ of the effective potential, as $V_\mathrm{PT}\sim \exp\left(-2\kappa r_*\right)$.
Different from Eq.~\eqref{asTruWKB}, the asymptotic QMNs line up in a primarily straight line in the frequency space whose angle to the real axis gets smaller as the perturbation moves away from the compact object.

Notably, in both cases above, the asymptotic QNMs exhibit an equally spaced distribution along the real frequency axis, irrespective of the proximity or magnitude of the discontinuity relative to the horizon.  
From a broader perspective on spectral instability, Jaramillo {\it et al.}~\cite{agr-qnm-instability-07, agr-qnm-instability-13} further investigated these implications by analyzing deterministic and random perturbations of the effective potential through pseudospectrum techniques in black hole perturbation theory. 
Their findings also reveal a migration of the pseudospectrum boundary toward the real frequency axis, reinforcing the concept of a universal instability in high-overtone modes triggered by ultraviolet (i.e., small-scale)  perturbations.
The exploration of spectral instability and its consequences hold potential implications for observational astrophysics, particularly in the realm of black hole spectroscopy~\cite{agr-BH-spectroscopy-review-04}.
In realistic astrophysical scenarios, gravitational wave sources rarely exist in isolation, instead interacting with surrounding matter and energy.
These interactions induce deviations from idealized symmetric metrics, pontentially causing emitted gravitational waves to diverge significantly from predictions based on pristine, isolated compact objects.
Such phenomena have sparked investigations into ``dirty'' black holes~\cite{agr-bh-thermodynamics-12, agr-qnm-33, agr-qnm-34, agr-qnm-54}, opening new frontiers in black hole perturbation theory.
The asymptotic modes aligning near the real axis are intricately linked to the concept of echoes, late-stage ringing waveforms first proposed by Cardoso {\it et al.}~\cite{agr-qnm-echoes-01, agr-qnm-echoes-review-01}.
As potential observables, echoes may serve as discriminators between similar gravitational systems through their distinct near-horizon properties.
This concept has catalyzed extensive research into echoes across various systems, including exotic compact objects like gravastars~\cite{agr-eco-gravastar-02, agr-eco-gravastar-03}, boson stars~\cite{agr-eco-gravastar-07}, and wormholes~\cite{agr-wormhole-01, agr-wormhole-02, agr-wormhole-10, agr-wormhole-11}.
Mark {\it et al.}~\cite{agr-qnm-echoes-15} have investigated echoes via the properties of the Green's function, which have been explored further by examining the function's asymptotic poles~\cite{agr-qnm-echoes-20, agr-qnm-echoes-45, agr-qnm-instability-65}.
In studies of Damour-Solodukhin wormholes~\cite{agr-wormhole-12}, Bueno {\it et al.}~\cite{agr-qnm-echoes-16} investigated echoes by solving for specific frequencies where the transition matrix becomes singular, illuminating the intricate relationship between spacetime geometry and QNMs.
The interplay between spectral instability, echoes, and causality has been a focal point of recent research~\cite{agr-qnm-instability-08, agr-qnm-instability-13, agr-qnm-instability-14, agr-qnm-instability-15, agr-qnm-instability-16, agr-qnm-instability-18, agr-qnm-instability-19, agr-qnm-instability-26, agr-qnm-echoes-22, agr-qnm-echoes-29, agr-qnm-echoes-30, agr-qnm-instability-23, agr-qnm-instability-29, agr-qnm-instability-32, agr-qnm-instability-33, agr-qnm-instability-43, agr-qnm-echoes-35}.
Notably, Cheung {\it et al.}~\cite{agr-qnm-instability-15} demonstrated that even fundamental modes can be destabilized under generic perturbations.
Recent studies have further scrutinized the physical significance of such instabilities~\cite{agr-qnm-instability-55, agr-qnm-instability-56, agr-qnm-instability-57, agr-qnm-instability-58}, and in particular, they have been attributed to an interplay between the asymptotic behavior of the Green's function near the singularity associated with a QNM and the spatial translation applied to the effective potential's perturbation~\cite{agr-qnm-instability-55}.
While high overtones typically reside farther from the real frequency axis, their collective effect might still manifest itself in phenomena such as echoes~\cite{agr-qnm-echoes-15, agr-qnm-echoes-16, agr-qnm-echoes-20, agr-qnm-echoes-45, agr-qnm-instability-65} with possible observational consequences.
As a result, the precise evaluation of QNMs in metrics subject to spectral instability remains a crucial endeavor in advancing our understanding of these complex astrophysical systems.

To date, Leaver's continued fraction method~\cite{agr-qnm-continued-fraction-01, agr-qnm-continued-fraction-04, agr-qnm-continued-fraction-12, agr-qnm-continued-fraction-23} remains unparalleled in its precision for calculating QNMs.
Nollert's extension of this algorithm has enabled the evaluation of the high overtones~\cite{agr-qnm-continued-fraction-12}.
Regarding the asymptotic QNMs of the Schwarzschild black holes, Nollert~\cite{agr-qnm-continued-fraction-12} and Motl's~\cite{agr-qnm-continued-fraction-23} studies showed that 
\bqn
\omega_n \sim \frac{\ln 3}{8\pi M}- i \frac{(2n+1)\pi}{8\pi M} + O\left(n^{-\frac12}\right) ,\label{asSch}
\eqn
where the high overtones asymptotically approach a line parallel to the imaginary frequency axis, drastically in contrast with the behavior given by Eqs.~\eqref{asTruWKB} and~\eqref{asTruWKB2}.
The matrix method~\cite{agr-qnm-lq-matrix-01,agr-qnm-lq-matrix-02,agr-qnm-lq-matrix-03,agr-qnm-lq-matrix-04}, developed by some of us, reformulates the QNM problem into a matrix equation for complex frequencies.
While in some ways reminiscent of the continued fraction method, it differs primarily in expanding the wave function over discrete coordinate grids.

When applied to effective potentials with discontinuities, most approaches for evaluating black hole QNMs face limitations.
For example, standard WKB formulae~\cite{agr-qnm-WKB-01, agr-qnm-WKB-02, agr-qnm-WKB-03, agr-qnm-WKB-05} calculate quasinormal frequencies using only the values of the effective potential and its derivatives at the maximum, disregarding any perturbations located elsewhere.
The monodromy method~\cite{agr-qnm-40} is known for its effectiveness in assessing high overtones.
However, it is based on analytic continuation of the wave function in coordinate space, and encounters complications due to discontinuities.

In the literature, a few approaches have been developed for metrics with discontinuities.
Kokkotas and Schutz~\cite{agr-qnm-star-07} employed numerical integration to study QNMs in pulsating relativistic stars with surface discontinuities.
Leins {\it et al.}~\cite{agr-qnm-star-08} modified the continued fraction method, which was adopted in subsequent studies~\cite{agr-qnm-34, agr-qnm-star-35, agr-qnm-54}.
The matrix method was also adapted to handle metrics with discontinuities~\cite{agr-qnm-lq-matrix-06}.
Comparative analysis demonstrated that this modified matrix method accurately solves for low-lying modes.
However, the existing recipes are largely applicable only to the first few low-lying modes.
Given the significance of spectral instability, precise evaluation of QNMs for effective potentials with discontinuities, especially high overtones, remains a crucial area of investigation.

The present study is motivated by the above considerations.
We extend Leaver's continued fraction method to evaluate black hole QNMs in systems where the effective potential exhibits a discontinuity.
We are particularly interested in high overtones, which are significantly altered due to spectral instability.
The generalized algorithm is characterized by two key elements.
First, the expansion of the wavefunction is not necessarily carried out around the horizon; specifically, it can be performed at the point of discontinuity.
Second, the recurrence relations between expansion coefficients have to be modified to incorporate the Israel-Lanczos-Sen junction conditions~\cite{agr-collapse-thin-shell-03}.

The remainder of the paper is organized as follows.
In the following section, we elaborate on an extended continued fraction method, specifically aimed at high overtones for black hole metrics exhibiting discontinuity.
The key feature of the proposed approach resides in the choice of expansion point and implementation of the junction condition.
In Sec.~\ref{sec3}, we apply this algorithm to evaluate high overtones of a few different types of modified Regge-Wheeler effective potentials containing discontinuity.
For the low-lying modes, the numerical results are compared to those obtained using the matrix method and applying the Prony method to the ringdown waveform constructed using the finite difference method.
Satisfactory agreement is achieved.
Also, the asymptotic behavior of the high overtones is analyzed, and consistency is ensured by employing different asymptotic wavefunctions and grid sizes.
Sec.~\ref{sec4} includes further discussions and concluding remarks.
We relegate the somewhat tedious analytical derivations to the appendices.

\section{Review of the continued fraction method}\label{sec2}

In this section, we first briefly revisit the formalism for the continued fraction method and its application to high overtones.
We then generalize the approach to metrics containing a discontinuity.
The present study primarily resides in the following theoretical setup in which the study of black hole perturbations can be simplified by exploring the radial part of the master equation~\cite{agr-qnm-review-03,agr-qnm-review-06},
\begin{eqnarray}
\frac{\partial^2}{\partial t^2}\Psi(t, r_*)+\left(-\frac{\partial^2}{\partial r_*^2}+V_\mathrm{eff}\right)\Psi(t, r_*)=0 .
\label{master_eq_ns}
\end{eqnarray}
The effective potential $V_\mathrm{eff}$ is determined by the given spacetime metric, spin, and angular momentum of the perturbation.
The Regge-Wheeler potential for a perturbation of spin $s$ and angular momentum $\ell$ in the Schwarzschild black hole with mass $M$ reads
\bqn
V_\mathrm{eff}=V_\mathrm{RW}(r;r_h)=f\left[\frac{\ell(\ell+1)}{r^2}+(1-{s}^2)\frac{r_h}{r^3}\right],
\label{V_RW}
\eqn
where the metric function
\bqn
f=1-r_h/r ,
\label{f_master}
\eqn
and the horizon $r_h=2M$.
In Eq.~\eqref{master_eq_ns}, $r_*\in (-\infty, +\infty)$ is the tortoise coordinate, which is related to the radial coordinate $r$ according to $dr_* = dr/f$. 

The QNMs, $\omega$, can be obtained by evaluating the zeros of the Wronskian
\begin{eqnarray}
W(\omega)\equiv W(g, h)=g(\omega,r_*)h'(\omega, r_*)-h(\omega,r_*)g'(\omega,r_*) ,
\label{pt_Wronskian}
\end{eqnarray}
where $'\equiv d/dr_*$, and $g$ and $h$ are the solutions of the corresponding homogeneous equation of Eq.~\eqref{master_eq_ns} in frequency domain~\cite{agr-qnm-review-02},
\begin{eqnarray}
\left[-\omega^2-\frac{d^2}{dr_*^2}+V_\mathrm{eff}\right]\widetilde{\Psi}(\omega, r_*)=0 ,
\label{pt_homo_eq}
\end{eqnarray}
with appropriate boundary conditions, namely,
\begin{eqnarray}
\begin{array}{cc}
g(\omega, r_*)\sim e^{-i\omega r_*}    &  r_*\to -\infty  ,\cr\\
h(\omega, r_*)\sim e^{i\omega r_*}     &  r_*\to +\infty   .
\end{array} 
\label{pt_boundary}
\end{eqnarray}

Following Leaver's recipe for the Regge-Wheeler potential, one factors the wavefunction $\widetilde{\Psi}$ by explicitly considering its asymptotic behavior at the horizon and spatial infinity, namely,
\bqn
&\widetilde{\Psi}\underset{r\to r_h}{\rightarrow} \left(\frac{r-r_h}{r_h}\right)^{-i\omega r_h} ,\\
&\widetilde{\Psi}\underset{r\to +\infty}{\rightarrow} \left(\frac{r}{r_h}\right)^{i\omega r_h}e^{i\omega r},
\eqn
so that we have
\bqn
\widetilde{\Psi} = \left(\frac{r-r_h}{r_h}\right)^{-i\omega r_h} \left(\frac{r}{r_h}\right)^{2i\omega r_h} e^{i\omega(r-r_h)} R(z) 
,\lb{CFPsiDef}
\eqn
where
\bqn
R(z)=\sum_{n=0}^{\infty} a_n z^n .
\eqn
One Taylor expands the transformed wave function $R(z)$ in terms of $z \equiv \frac{r-r_h}{r}\in (0, 1)$.
By substituting the expansion Eq.~\eqref{CFPsiDef} into the master equation Eq.~\eqref{master_eq_ns} for the Regge-Wheeler potential Eq.~\eqref{V_RW}, one finds the following three-term recurrence relation~\cite{agr-qnm-continued-fraction-01}
\bqn
\alpha_1 a_1 + \beta_1 a_0 &=& 0, \nb\\
\alpha_n a_{n} +\beta_{n} a_{n-1} +\gamma_n a_{n-2} &=& 0, \ \ \ n=2, 3,\cdots.
\lb{threerecu}
\eqn
In particular, for $r_h=1$, the coefficients $\alpha_n, \beta_n$, and $\gamma_n$ are given by
\bqn
\alpha_n &=& n^2+(-2i\omega+2)n-2i\omega +1, \nb\\
\beta_n &=& -2n^2+(8i\omega-2)n+8\omega^2+4i\omega-\ell(\ell+1))+s^2-1 ,\nb\\
\gamma_n &=& n^2-4i\omega n -4\omega^2 -s^2 .
\lb{threerecu1}
\eqn
The first line of Eq.\eqref{threerecu} gives
\bqn
\frac{a_1}{a_0} = -\frac{\beta_1}{\alpha_1} ,
\eqn
and the second line of  Eq.\eqref{threerecu} can be used to derive the continued fraction
\bqn
\frac{a_1}{a_0} = \frac{-\gamma_2}{\beta_2 -\dfrac{\alpha_2 \gamma_3}{\beta_3 -\dfrac{\alpha_3 \gamma_4}{\beta_4 - \cdots}}} .
\lb{a1overa0}
\eqn
Putting these together leads to the equation
\bqn
\beta_1
 = \frac{\alpha_1 \gamma_2}{\beta_2 -\dfrac{\alpha_2 \gamma_3}{\beta_3 -\dfrac{\alpha_3 \gamma_4}{\beta_4 - \cdots}}} .
 \lb{contfraceqn}
\eqn
By truncating this continued fraction at a sufficiently large depth, one obtains an equation for $\omega$ that can be solved using standard root-finding techniques.  
It turns out that to find more than just a few roots of (\ref{contfraceqn}) we need to ``invert'' the equation.  Usually the $n^{\text{th}}$ QNM, $\omega_n$, is most easily found using the $n^{\text{th}}$ inversion which is defined as 
\bqn
\beta_n - \frac{\alpha_{n-1} \gamma_n}{\beta_{n-1} -} \frac{\alpha_{n-2} \gamma_{n-1}}{\beta_{n-2} -}\cdots \frac{\alpha_1\gamma_2}{\beta_1}
=
\frac{\alpha_{n} \gamma_{n+1}}{\beta_{n+1} -} \frac{\alpha_{n+1} \gamma_{n+2}}{\beta_{n+2} -} \frac{\alpha_{n+2}\gamma_{n+3}}{\beta_{n+3} -}\cdots.
\label{eqInv}
\eqn
Nollert \cite{agr-qnm-continued-fraction-12} refined this technique by estimating the remainder when the continued fraction is truncated.  

Besides the orginal radial and tortoise coordinates, we also employ hyperboloidal coordinates~\cite{agr-qnm-hyperboloidal-02, agr-qnm-hyperboloidal-03, agr-qnm-instability-07}.
Specifically, one first transforms the coordinates into dimensionless quantities $(\overline{t}, \overline{x})$ by introducing a length scale $\lambda$ and set 
\begin{eqnarray}
\overline{t}=\frac{t}{\lambda},~~~~\overline{x}=\frac{r_{*}}{\lambda},~~~~{\hat{V}_\mathrm{eff}=\lambda^2 V_\mathrm{eff}},
\label{dimensionless_quantities}
\end{eqnarray}
where the choice of $\lambda$ is rather arbitrary, and does not affect the resultant master equation in the present case. 
Subsequently, the compactified hyperboloidal coordinates $(\tau,x)$ are defined by
\begin{eqnarray}
\overline{t}&=&\tau+H(\overline{x}),\nb\\
\overline{x}&=&G(x),
\label{compactified_hyperboloidal_approach}
\end{eqnarray}
where the function $G$ introduces a spatial compactification, while the height function $H$ is defined to guarantee that the boundary is the future event horizon or null infinity.
For a Schwarzschild black hole, we adopt~\cite{agr-qnm-instability-07}
\begin{eqnarray}
H(\sigma) &=& \frac12\left(\frac{1}{\sigma} -\ln(1-\sigma)- \ln \sigma\right) , \nb\\
G(\sigma) &=& \frac12\left(\frac{1}{\sigma}+\ln(1-\sigma)-\ln\sigma\right) ,\label{HeightCompac}
\end{eqnarray}
with $\lambda=4M=2r_h$, $\sigma=2M/r=r_h/r$.
The boundary is defined at $\sigma=1$ for the horizon and $\sigma=0$ for spatial infinity.
By rewriting the wavefunction in $(\tau, x)$ and using the separation of variables
\begin{eqnarray}
\Psi(\tau,x)=e^{-i\omega \lambda \tau}\phi(x)=e^{-i\omega t}e^{i\omega \lambda H(\overline{x})}\phi(x),
\label{wave_function}
\end{eqnarray}
where $x=1-\sigma=(r-r_h)/r\in (0,1)$, one finds the Fourier transform of Eq.~\eqref{master_eq_ns} in hyperboloidal time $\tau$, which takes the form 
\begin{eqnarray}
&&(x-1)^2x \phi''(x)+(1+x^2(3-4i\omega)+x(-4+8i\omega)-2i\omega)\phi'(x)\nb\\
&&-(1+\ell+\ell^2+s^2(x-1)-x-4i\omega+4ix\omega-8\omega^2+4x\omega^2)\phi(x) = 0 .
\lb{CFModPsiDef0}
\end{eqnarray}
Instead of Eq.~\eqref{CFPsiDef}, we now consider the expansion
\bqn
\phi(x) = \sum_{n=0}^{\infty} a_n x^n ,
\lb{CFModPsiDef}
\eqn
and it is readily verified that this leads to precisely the same three-term recurrence relation, namely, Eqs.~\eqref{threerecu} and~\eqref{threerecu1}.
In other words, Eqs.~\eqref{CFPsiDef} and~\eqref{CFModPsiDef} have precisely the same expansion coefficients $a_n$.
The reason behind this seeming coincidence is that
the asymptotic form of the wavefunction for the original Regge-Wheeler potential is taken care of by the factor $e^{i \omega \lambda H(\bar{x})}$ in Eq.~\eqref{wave_function}. More specifically, Eq.~\eqref{CFPsiDef} is identical, up to a constant, to the spatial part of Eq.~\eqref{wave_function} in the $r$ coordinate.
Therefore, as noted in~\cite{agr-qnm-hyperboloidal-22}, the hyperboloidal approach is equivalent to Leaver's continued fraction method. 
Therefore, it does not constitute a mathematically independent approach. 
Further discussion of the hyperboloidal approach is left to Appx.~\ref{app2}.

\section{Extended continued fraction method for a Regge-Wheeler potential with a jump discontinuity}\label{sec3}

In this study, we consider the following modified Regge-Wheeler effective potential that contains a small step, implemented by a discontinuity at $r_c$: 
\bqn
V_\mathrm{eff} 
= \begin{cases}
   V_\mathrm{RW}(r; r^-_h), &  r \le r_{c}, \\
   V_\mathrm{RW}(r; r^+_h), &  r > r_{c}, 
\end{cases}
\lb{Veff_MRW}
\eqn
where $r^-_h=2M$ and $r^+_h=2(M+\delta M)$.
It corresponds to a physically relevant scenario where a mass shell $\delta M$ is located at $r = r_c$ surrounding a Schwarzschild black hole.
We can compare this discontinuous potential with a simpler one that is truncated at $r_c$:
\bqn
V_{\mathrm{trunc}} 
= \begin{cases}
   V_\mathrm{RW}(r; r^-_h), &  r \le r_{c}, \\
   0, &  r > r_{c} .
\end{cases} 
\lb{Veff_MRWTrctd2}
\eqn


The discontinuity introduced in Eqs.~\eqref{Veff_MRW} and \eqref{Veff_MRWTrctd2} is governed by the Israel-Lanczos-Sen's junction condition~\cite{agr-collapse-thin-shell-03, book-general-relativity-Poisson}.
We have two different wave equations on each side of $r_c$.  The solutions to these wave equations must satisfy the condition~\cite{agr-qnm-34}:
\bqn
\lb{Israel}
\lim_{\epsilon\to 0^+}\left[\frac{\widetilde{\Psi}'(r_*^c +\epsilon)}{\widetilde{\Psi}(r_*^c +\epsilon)}-\frac{\widetilde{\Psi}'(r_*^c -\epsilon)}{\widetilde{\Psi}(r_*^c -\epsilon)}\right]=\kappa \ ,
\label{eq:junctioncondition}
\eqn
where the prime represents the derivative with respect to the tortoise coordinate $r_*$
and $r_*^c$ is the value of the tortoise coordinate at $r_c$.
For the Schr\"odinger-type master equation Eq.~\eqref{pt_homo_eq},
\bqn
\lb{kappa}
\kappa = \lim_{\epsilon\to 0^+}\int_{r_*^c-\epsilon}^{r_*^c+\epsilon} V_{\mathrm{eff}}(r_*)dr_*~.
\eqn
If one considers a moderate finite jump, then $\kappa =0$ and the above relation simplifies to the condition of a vanishing Wronskian
\bqn
\lb{Wronskian}
W(\omega) \equiv \widetilde{\Psi}(r_*^c+\epsilon)\widetilde{\Psi}'(r_*^c-\epsilon) - \widetilde{\Psi}(r_*^c-\epsilon)\widetilde{\Psi}'(r_*^c+\epsilon) = 0~.\label{WronskianZero}
\eqn

It has been shown by Leins {\it et al.}~\cite{agr-qnm-star-08}, in the context of neutron stars, and Onazawa {\it et al.}~\cite{agr-qnm-continued-fraction-19}, in the context of extremal black holes, that to apply the continued fraction method, one does not need to expand the wavefunction in Eq.~\eqref{CFPsiDef} around the horizon $r=r_h$, which potentially becomes an irregular singular point.  
In our case, the most natural choice is to expand about $r_c$, where the two solutions are matched via a junction condition.

It is noted that, whether one uses the tortoise coordinate $r_*$ or the hyperboloidal coordinate $x$, a change occurs at the coordinate value corresponding to $r_c$.
This is because both coordinates depend on the mass, which changes beyond the discontinuity.  
In the case of the tortoise coordinate, we have
\begin{eqnarray}
r^-_* &=&    r + r^-_h \ln{\left(\frac{r}{r^-_h}-1\right)} ~~~~~~~~~~~~~~~  r \le r_{c}, \nonumber \\
 r^+_* &=& r + r^+_h \ln\left(\frac{r}{r^+_h}-1\right) + C ~~~~~~~~  r > r_{c} ,
\end{eqnarray}
where we have the freedom to choose the constant $C$ so the two coordinates agree at $r=r_c$.
Note that the junction condition, Eq.~\eqref{WronskianZero}, is derived by integrating the master equation, Eq.~\eqref{pt_homo_eq}, with respect to the tortoise coordinate, so the derivative of the wavefunction is with respect to $r_*^-$ for $r\le r_c$ and $r_*^+$ for $r > r_c$.

In place of Eq.~\eqref{CFPsiDef}, we consider the asymptotic wavefunction
\bqn
\widetilde{\Psi} = 
\begin{cases}
    \psi^-_{\rm{asp}}  R^-(r) & r\le r_c \\
    \psi^+_{\rm{asp}}  R^+(r) & r > r_c
\end{cases} ,
\lb{CFPsiDef2}
\eqn
where 
\bqn
\psi^\pm_{\rm{asp}} &=& \left(\frac{r-r^\pm_h}{r^\pm_h}\right)^{\pm i\omega r^\pm_h}  e^{\pm i\omega r} ,\nonumber \\
R^\pm(r) &=& \sum_{n=0}^{\infty} \ a^\pm_n z_\pm^n ,
\lb{Veff_MRWTrctdAA}
\eqn
and
\bqn
z_- &=& \frac{r-r_c}{r+k} ,\nonumber \\ 
z_+ &=& \frac{r-r_c}{r-r^+_h} .
\lb{Veff_MRWTrctdAA1}
\eqn
The constant $k$ is a large\footnote{In our calculations we find $k=1000r^-_h$ is sufficiently large.} real number introduced to ensure the convergence of the expansion.  
Note that if one were to take $k=0$ then for $r_c>2r^-_h$ the argument of the expansion $|r-r_c|/r$ becomes larger than $1$ at $r=r^-_h$.
To keep the series convergent without $k$, one would have to add an extra condition on the coefficients $a_n$, which we would like to avoid.
For $z_+$, it is useful to include the term $r^+_h$ in the denominator as it reduces the number of terms that appear in the recurrence relation.  
It turns out that for $r_c \ge 2$ we can replace $r^+_h$ with $0$ and still get the same results.  
But, for $r_c < 2$, when replacing $r^+_h$ with $0$, one encounters numerical difficulties.    

By substituting the above expansion into the master equation~\eqref{pt_homo_eq}, both expansions lead to either a five (for $r > r_c$) or six-term (for $r \le r_c$) recurrence relation. 
For $r\le r_c$ and $r_h^- = 2M$, the recurrence relation has the form
\bqn
\alpha^-_1 a^-_1 + \beta^-_1 a^-_0 &=& 0, \nb\\
\alpha^-_2 a^-_2 + \beta^-_2 a^-_1 + \gamma^-_2 a^-_0 &=& 0, \nb\\
\alpha^-_3 a^-_{3} +\beta^-_3 a^-_2 +\gamma^-_3 a^-_{1} +\delta^-_3a^-_{0} &=& 0, \nb\\
\alpha^-_4 a^-_{4} +\beta^-_4 a^-_3 +\gamma^-_4 a^-_{2} +\delta^-_4a^-_{1}+ \epsilon^-_4 a^-_{0} &=& 0, \nb\\
\alpha^-_n a^-_{n} +\beta^-_n a^-_{n-1} +\gamma^-_n a^-_{n-2} +\delta^-_na^-_{n-3}+ \epsilon^-_na^-_{n-4}+ \zeta^-_na^-_{n-5} &=& 0, \ \ \ n=5,6,\cdots,\lb{SixTermHBtruncAA}
\eqn
with coefficients
\bqn
\alpha^-_n &=& n r_c^2(n-1) (r^-_h-r_c) ,\nb\\
\beta^-_n &=& r_c (n-1) \left[k\left(-3 (n-2) r_c+(2 n-5) r^-_h+2 i \omega r_c^2\right)+r_c \left(2 (n-1) r_c-3 (n-1) r^-_h+2 i \omega  r_c^2\right)\right] ,\nb \\
\gamma^-_n&=&k^2 \left[r_c \left(\ell(\ell+1)-3
(n^2-5 n+6)\right)+(n^2-6 n+5) r^-_h+6 i (n-2) \omega  r_c^2\right] \nb\\
&&+k r_c \left[2 \ell(\ell+1) r_c +3 (n-2)(r^-_h+2 i \omega  r_c^2)+6 (n-2)^2
(r_c-r^-_h)-6 r^-_h\right]\nb\\
&&+r_c^2 \left[r_c(\ell(\ell+1)-n^2+3 n-2) +3 (n^2-3 n+1) r^-_h\right] ,\nb\\
\delta^-_n &=& k^3 \left[\ell(\ell+1)-(n-3) (n-6 i \omega   r_c-4)\right]-(n-4) n r_c^2 r^-_h \nb\\
&&+k^2 \left[2 \ell(\ell+1) r_c+3\left((n-3)
(r^-_h+2 i \omega r_c^2)+(n-3)^2 (2 r_c-r^-_h)+r^-_h\right)\right]\nb\\
&&+k r_c \left[r_c
\left(\ell(\ell+1)-3 (n^2-5 n+6)\right)+3 (2 n^2-11 n+17)
r^-_h\right] ,\nb\\
\epsilon^-_n &=& -k (n-4) (-2 i k^3 \omega
+k^2 (-2 n-2 i \omega r_c+8)+3 k (n-3) r_c-3 k (n-4)
r^-_h+(2 n-5) r_c r^-_h) ,\nb \\
\zeta^-_n &=& -k^2 (n-5) (n-4) (k+r^-_h) .
\lb{SixTermLHSCoefficients}
\eqn

Similarly, for $r>r_c$  one finds a five-term recurrence relation in $a^+_n$ with the following coefficients,
\bqn
\alpha^+_n &=& (n-1) n r_c^2 ,\nb \\
\beta^+_n &=&  r_c (1 - n) \left[-3 r^+_h + 2 (n-1) (r_c + r^+_h) -  2 i r_c^2 \omega\right] ,\nb \\
\gamma^+_n &=& r_c r^+_h (23 + \ell(\ell+1) - 18 n + 4 n^2)  +  {r^+_h}^2 (5 - 6 n + n^2) \nb \\
&&+ r_c^2 \left[-\ell(\ell+1) + (n-2) (n-1 - 6 i r^+_h\omega)\right] ,\nb \\
\delta^+_n &=& -r^+_h \left[ r^+_h\left(21 + \ell(\ell+1) - 14 n +
        2 n^2\right) + r_c \left(-\ell(\ell+1) + 2 (n-3)^2 + n - 6 i (n-3) r^+_h \omega\right)\right] ,\nb \\
\epsilon^+_n &=& {r^+_h}^2(n-4)  (n-4 - 2 i r^+_h \omega) .
\lb{FiveTermRHSCoefficients}
\eqn

In this case, the lowest-order recurrence relation becomes trivial because $\alpha^\pm_1=\beta^\pm_1=0$.
Therefore, unlike Eq.~\eqref{threerecu}, the first line of Eq.~\eqref{SixTermHBtruncAA} does not determine the ratio between $a_0$ and $a_1$.  
However, the missing constraint is naturally supplied by the junction condition, Eq.~\eqref{WronskianZero}, ensuring that the recurrence relations remain neither under- nor overdetermined.

To be more specific, for the truncated effective potential $V_\text{trunc}$ in Eq.~\eqref{Veff_MRWTrctd2}, we have $\psi^+_\text{asp} R^+(r) =e^{i \omega r^+_*}$, and the junction condition Eq.~\eqref{Wronskian} becomes
\bqn
\frac{a^-_1}{a^-_0} \left.\frac{dz_-}{dr^-_*}\right|_{r^c_*} +\left. \frac{1}{\psi^-_{\rm{asp}}} \frac{d\psi^-_{\rm{asp}}}{dr^-_*}\right|_{r^c_*} = i\omega,\lb{WronskianxcRHS}
\eqn
which can be solved for the ratio $a^-_1/a^-_0$.  
We then get a continued fraction equation by using this for the l.h.s. of Eq.~\eqref{a1overa0}.
Note that in this case, we can use $r^+_* = r^-_*$.
For the more general case of effective potential Eq.~\eqref{Veff_MRW}, we have
\bqn
\frac{a^-_1}{a^-_0} \left.\frac{dz_-}{dr^-_*}\right|_{r^c_*} +\left. \frac{1}{\psi^-_{\rm{asp}}} \frac{d\psi^-_{\rm{asp}}}{dr^-_*}\right|_{r^c_*} = 
\frac{a^+_1}{a^+_0} \left.\frac{dz_+}{dr^+_*}\right|_{r^c_*} +\left. \frac{1}{\psi^+_{\rm{asp}}} \frac{d\psi^+_{\rm{asp}}}{dr^+_*}\right|_{r^c_*}.
\lb{Wronskianxc}
\eqn
Again using Eq.~\eqref{a1overa0}, the above gives an equation involving two continued fractions, one for $a_1^-/a_0^-$ and the other for $a_1^+/a_0^+$.  This equation can be solved for $\omega$ by truncating each continued fraction at a sufficient depth.

\section{The matrix method and its extension for a Regge-Wheeler potential with a jump discontinuity}

In this section, we summarize the implementation of the matrix method for black hole QNMs and its extension to effective potentials with discontinuities. 
For further technical details, we refer the readers to the original work~\cite{agr-qnm-lq-matrix-06}.

Our starting point is the standard master equation~\eqref{pt_homo_eq} expressed in the tortoise coordinate, which is then mapped to a compact coordinate $z\in[0,1]$ so that the inner and outer boundaries lie at $z=0$ and $z=1$.
Again, the known asymptotic behaviour at the boundaries is factored out, and the remaining wavefunction $R(z)$ is manifestly regular in the whole interval.  
In terms of $R(z)$ the differential equation becomes $\mathcal{H}(\omega,z)R(z)=0$, where $\mathcal{H}$ is a second-order linear differential operator in $z$ that depends on $V_{\mathrm{eff}}$ and the chosen field redefinition.  
It is convenient to define $F(z)=z(1-z)R(z)$ so that the quasinormal boundary conditions reduce to $F(0)=F(1)=0$. With this substitution, the governing equation can be expressed schematically as $\mathcal{G}(\omega,z)F(z)=0$, where $\mathcal{G}$ is another linear operator obtained from $\mathcal{H}$.  

Subsequently, the equation is discretized by introducing $N$ grid points $\{z_i\}$, such as Chebyshev grid, in the interval $[0,1]$, and defining $f_i\equiv F(z_i)$ and $\mathcal{F}=(f_1,\dots,f_N)^T$.  
Derivative terms in $\mathcal{G}$ are represented by local interpolation at each grid point so that the action of $\mathcal{G}$ becomes an algebraic system 
\bqn
\overline{\mathcal{M}}(\omega)\,\mathcal{F}=0 , \label{algeMM}
\eqn
where $\overline{\mathcal{M}}(\omega)$ is an $N\times N$ matrix whose elements depend only on the grid and on the effective potential $V_{\mathrm{eff}}$.  
To enforce the boundary conditions, the first and last rows of $\overline{\mathcal{M}}(\omega)$ are replaced by $f_1=0$ and $f_N=0$, which leads to the modified matrix $\mathcal{M}(\omega)$ defined by 
\[
\mathcal{M}_{ki}=\delta_{1i}\ \text{for}\ k=1,\quad \mathcal{M}_{ki}=\overline{\mathcal{M}}_{ki}\ \text{for}\ k=2,\dots,N-1,\quad \mathcal{M}_{ki}=\delta_{Ni}\ \text{for}\ k=N.
\]
The matrix equation Eq.~\eqref{algeMM} admits non-trivial solutions only when 
\bqn
\det\overline{\mathcal{M}}(\omega)=0 , \label{MMDetEq}
\eqn
and the roots of this determinant provide the quasinormal frequencies.  

When the effective potential contains a discontinuity at some $z=z_c$, the Taylor expansion used in the discretization cannot be continued through this point, and the values of the wavefunction on both sides must satisfy a junction condition (cf. Eq.~\eqref{eq:junctioncondition} for thin-shell-type configurations).
In the discretized system, the grid is partitioned into two smooth subdomains, and the matrix rows that would otherwise enforce the differential equation at $z=z_c$ are replaced by a discrete version of the matching condition involving grid values on both sides of the discontinuity.  
This procedure results in an almost block-diagonal matrix in which the two subdomains are linked only by the interface row (and, via interpolation, the corresponding column), so that the determinant condition Eq.~\eqref{MMDetEq} again yields the complete QNM spectrum.  
As a result, the presence of the discontinuity is fully encoded in the algebraic structure of $\overline{\mathcal{M}}(\omega)$, while the rest of the implementation remains the same as in the original algorithm.

\section{Numerical implementation and results}\label{sec4}

We are now in a position to present the numerical results.
The calculations are carried out using the extended continued fraction method in the radial coordinate, the matrix method in hyperboloidal coordinates, and the Prony method applied to the temporal profile obtained using the finite difference method.
The latter two approaches can only be used to obtain the first few low-lying modes, but as discussed below, they yield results that are consistent with those obtained by the continued fraction method.
Different grid sizes and expansion orders have yielded consistent results, ensuring the numerical scheme's precision.

\subsection{Numerical setup of the continued fraction method}\label{sec3.0}

For the continued fraction method, the numerical procedure is summarized as follows.  
We construct a grid of points in a region near the origin of the complex frequency plane, which serve as initial guesses for a root finder (M\"uller's method in this case) used to solve Eq.~(\ref{WronskianxcRHS}) or (\ref{Wronskianxc}) over several inversions.
Within each inversion, we used two different depths of the truncated continued fraction and retained roots that appear in both depths (we consider roots to be the same if the distance between them is less than $10^{-8}$). 
From those roots, we keep the ones that persist between inversions. 
With the first few QNMs in hand, we use the slope defined by the last two (most damped) QNMs to predict the location of the next QNM. 
We repeat the process until the desired number of roots is found.  

The computation time increases with the absolute value of $\omega$, as such modes require a much greater depth in the continued fraction and thus higher precision in each calculation.  
This makes them increasingly difficult to find.  
Also, as the truncation point moves away from the black hole, increasingly higher precision is demanded.

Note that when solving Eq.~(\ref{Wronskianxc}), there are two continued fractions.  
For each guess of $\omega$, we evaluate the continued fraction for $a_1^+/a_0^+$ to some depth and then find inversions for the continued fraction corresponding to $a_1^-/a_0^-$ using the same depth.  
As the number of inversions increases, we increase the depth.

\subsection{Validation using the matrix and Prony methods}\label{sec3.x}

To ascertain our results, two different approaches are employed to evaluate the first few low-lying modes.
The extended matrix method is used to evalute the fundamental mode and the first overtone for the truncated Regge-Wheeler potential Eq.~\eqref{Veff_MRWTrctd2} with $r_c=3/2$ and $r_c=2$.
The method is implemented on the Chebishev grid by using the junction condition Eq.~\eqref{WronskianxcRHS}.
The resulting algebraic equation~\eqref{MMDetEq} is solved using the {\it Mathematica} subroutine {\it Eigensystem}.

We present the following numerical results, in units of $r_h^-$, demonstrating the consistency between the continued fraction method and the matrix method. 
For $r_c = 3/2$ and grid size $N = 10$, the fundamental mode and the first overtone are found to be $\omega_0 = 0.5055899643 - 0.4218252318i$ and $\omega_1 = 0.3390503047 - 1.2957801244i$, respectively.  
Similarly, for $r_c = 2$, the fundamental mode and the first overtone are $\omega_0 = 0.7259510126 - 0.3551794085i$ and $\omega_1 = 0.8141641390 - 1.1089404285i$.  
With an increased grid size of $N = 50$, the corresponding values are $\omega_0 = 0.5055959719 - 0.4218216823i$ and $\omega_1 = 0.3393478026 - 1.2959620252i$ for $r_c = 3/2$, and $\omega_0 = 0.7259507136 - 0.3551795471i$ and $\omega_1 = 0.8141469346 - 1.1089266495i$ for $r_c = 2$.  
Compared with the results obtained using the continued fraction method shown in Tab.~\ref{Tab.1}, the agreement, better than six significant figures, is satisfactory.  
For higher overtones or larger values of $r_c$, this level of accuracy can be maintained, though it requires increasing the grid size to ensure sufficient numerical precision.

As a further check on our results, the low-lying modes are also extracted using the Prony method from the time-domain evolution obtained from the finite difference method.
To generate the ringdown waveform for the differential equation \eqref{master_eq_ns}, we take as initial conditions a Gaussian pulse located at $r_* = \bar{r}_*$ moving toward the potential,
\bqn
\Psi(0, r_*) = \mathcal{A}~ \text{exp}\left(-\frac{(r_*- \bar{r}_*)^2}{2\sigma^2}\right) \\
\left. \frac{\partial}{\partial t}\Psi\right|_{t=0} = - \frac{\partial}{\partial r_*}\Psi(0, r_*).
\eqn
We use the values $\mathcal{A} = 30, \bar{r}_* = -40$ and $ \sigma=1$.  
However, the QNMs obtained are not sensitive to these specific values.  
Letting $\Psi^{i}_j=\Psi(t_i, r_{*j})$, where $t_i$ and $r_{*j}$ are the discretized values of $t$ and $r_*$, the temporal evolution of the differential equation \eqref{master_eq_ns} is given by
\bqn
\Psi_j^{i+1} = -\Psi_j^{i-1} +\Psi_j^i \left( 2 - 2 \left( \frac{\Delta t}{\Delta r_*}
\right)^2 - V(r_*) (\Delta t)^2 \right) + \left(\Psi_{j+1}^i - \Psi_{j-1}^i\right)\left(\frac{\Delta t}{\Delta r_*}\right)^2.
\eqn
For a given $t$-value, the values for $\Psi(r_*)$ are calculated from the left and right boundaries moving toward $r_*^c$.  The value of $\Psi$ at $r_*^c$ is determined using the junction condition \eqref{eq:junctioncondition}, which can be written using a one-sided second-order approximation for the derivative
\bqn
\frac{3 \Psi^i_j-4\Psi^i_{j-1}+\Psi^i_{j-2}}{2\Delta r_*}=\frac{-3 \Psi^i_j+4\Psi^i_{j+1}-\Psi^i_{j+2}}{2\Delta r_*}.
\eqn
Solving for $\Psi^i_j$ gives the value of $\Psi$ at the $j$-index corresponding to $r_*^c$.

In Fig.~\ref{Ringdown}, we showcase the temporal evolution of the solution $\Psi$ for the effective potential Eq.~\eqref{Veff_MRW} with $r_c = 3/2$ and $ \delta M = 0.005$.
The calculations are carried out using the grid size $\Delta t= 0.01$ and $\Delta r_*= 0.02$.
For the Prony method, we used the data on the interval $110 \le t \le 130$.
Comparing the results obtained with the Prony method for $r_c = 3/2$ and $r_c = 2$ when $\delta M = 0.005$, we found agreement to five significant figures for the fundamental mode and to three significant figures for the first overtone in both cases.
For higher values of $r_c$, the agreement worsens due to the appearance of echo modes outside the main ringdown signal, for which the Prony method is ill-suited, as discussed in~\cite{agr-qnm-instability-65}.

\subsection{Truncated Regge-Wheeler effective potential}\label{sec3.1}

Our numerical results, expressed in units of $r_h^-$, for the truncated Regge-Wheeler potential Eq.~\eqref{Veff_MRWTrctd2} are shown in Tabs.~\ref{Tab.1},~\ref{Tab.2},~\ref{Tab.3},~\ref{Tab.4}, and Figs.~\ref{RW_trunc_QNM_zoom} and~\ref{RW_trunc_QNM}. 
Calculations of $n=2000$ modes is carried out for different truncations at $r_c = 3/2$, $7/4$, $2$, $3$, $5$, $10$, $15$, $25$, $35$, $50$, and $100$.
These results are summarized in Figs.~\ref{RW_trunc_QNM_zoom} and~\ref{RW_trunc_QNM}.
For the first few low-lying modes, the precision of the numerical values has been verified by comparing them with the results obtained using the matrix method in~\cite{agr-qnm-lq-matrix-12}, finding agreement to all decimal places.

\begin{figure}
    \centering
    \includegraphics[width=0.453\linewidth]{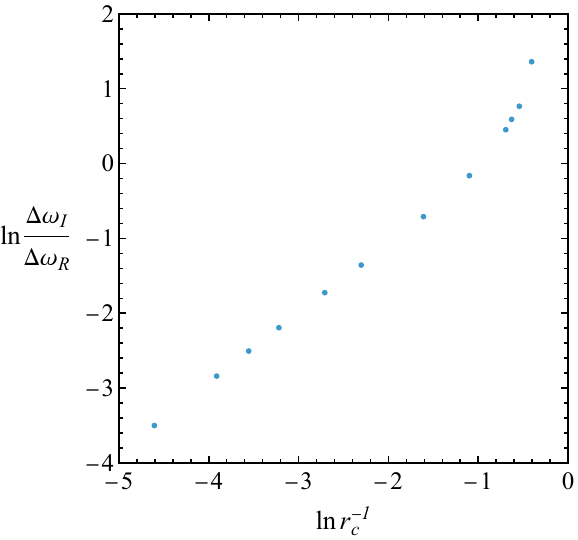}
    \includegraphics[width=0.4\linewidth]{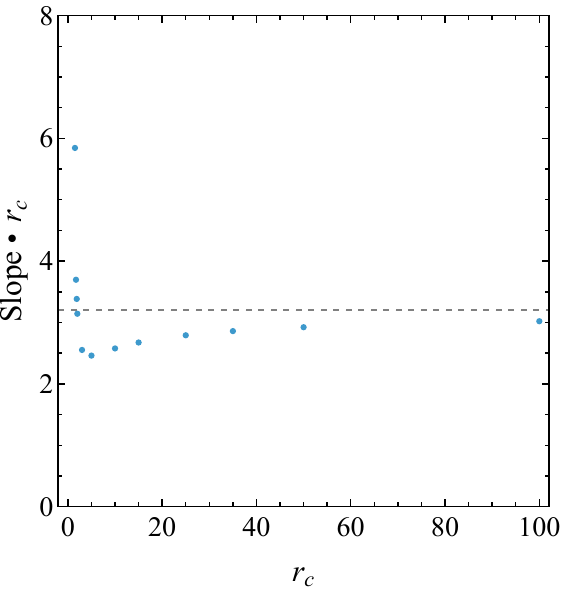}
    \caption{We show the slope of the QNM spectrum as a function of $1/r_c$ in logarithmic scale for the truncated Regge-Wheeler potential in order to make it easier to see the linear behavior as $r_c\rightarrow \infty$. In the right panel, we show the slope times $r_c$, as a function of $r_c$.  The dashed line located at $3.1$ is where the slope times $r_c$ asymptotes, consistent with Eq. \eqref{Jodinfit}.}
    \label{fig:asymp-slope}
\end{figure}

It is observed that as $r_c$ increases, the QNMs move away from the imaginary axis and tilt toward the real axis.  
At $r_c =100$, the QNMs are almost on the real axis.  
One observes that $\Delta \omega_R \propto 1/r_c$ and $\Delta \omega_I \propto 1/r_c^2$ for $r_c\gg r_h^-$, where $\omega_R$ and $\omega_I$ are the real and imaginary parts of $\omega$ respectively.  This is shown in Fig.~\ref{fig:asymp-slope}.
To be more precise, we found
\bqn
\Delta \omega_R &\approx & \frac{3.1}{r_c} ,\nb\\
\Delta \omega_I &\approx & -\frac{3.1 r_h^-}{r_c}\Delta\omega_R.\label{Jodinfit}
\eqn

Existing results indicate that the above asymptotic behavior of $\Delta \omega_R$ appears to be universal across all known examples of spectral instability~\cite{agr-qnm-instability-65}, in which the QNM spectrum tends to approach the real axis as $r_c$ increases~\cite{agr-qnm-instability-02, agr-qnm-instability-03, agr-qnm-instability-11, agr-qnm-lq-03}.  
Specifically, the asymptotic high overtones have been analytically evaluated for the truncated P\"oschl-Teller potential~\cite{agr-qnm-instability-65}, see Eq.~\eqref{asTruWKB2}, yielding  
\bqn
\Delta \omega_R &=& \frac{\pi}{r_c}, \nb\\
\Delta \omega_I &=& -\frac{\pi}{2\kappa r_c}\Delta\omega_R .
\eqn
This is, somewhat surprisingly, in good agreement with Eq.~\eqref{Jodinfit}.  
In fact, we speculate that with additional data (including higher overtones and larger values of $r_c$), the proportionality constant in Eq.~\eqref{Jodinfit} will approach $\pi$.

\subsection{Modified Regge-Wheeler potential with a jump discontinuity}\label{sec3.2}

Our numerical results for the modified Regge-Wheeler potential in Eq.~\eqref{Veff_MRW} with $\delta M = 0.001\,r_h^{-1}$ are presented, in units of $r_h^{-1}$, in Tabs.~\ref{Tab. 2.1},~\ref{Tab. 2.2},~\ref{Tab. 2.3},~\ref{Tab. 2.4}, and Figs.~\ref{RW_2Sided_QNM_zoom} and~\ref{RW_2Sided_QNM}.
Again, as $r_c$ increases, the QNMs move away from the imaginary axis and tilt toward the real axis, with  
\bqn
\Delta \omega_R &\approx& \frac{3.1}{r_c}, \nb\\
\Delta \omega_I &\approx& -\frac{3.1 r_h^-}{r_c}\Delta\omega_R. \label{Jodinfit2}
\eqn
for large values of $r_c$.  
By comparing Eqs.~\eqref{Jodinfit} and~\eqref{Jodinfit2}, one notes that, up to numerical uncertainty, the inclination of the high overtones in these two cases is identical.
This naturally raises a question of how the inclination of the asymptotic spectrum depends on the size of the mass shell $\delta M$.

\begin{figure}
    \centering
    \includegraphics[height=0.5\linewidth]{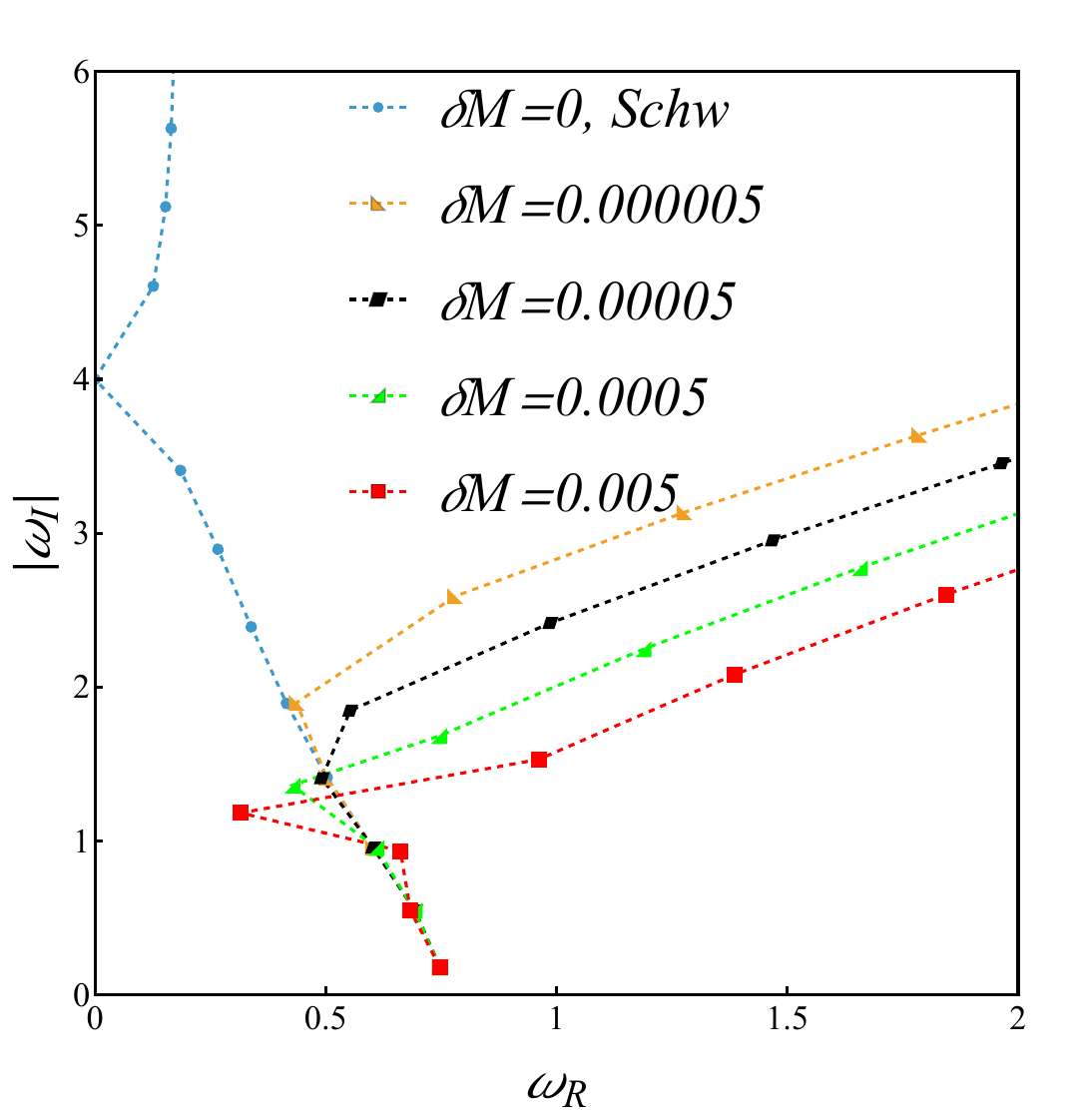}
     \includegraphics[height=0.5\linewidth]{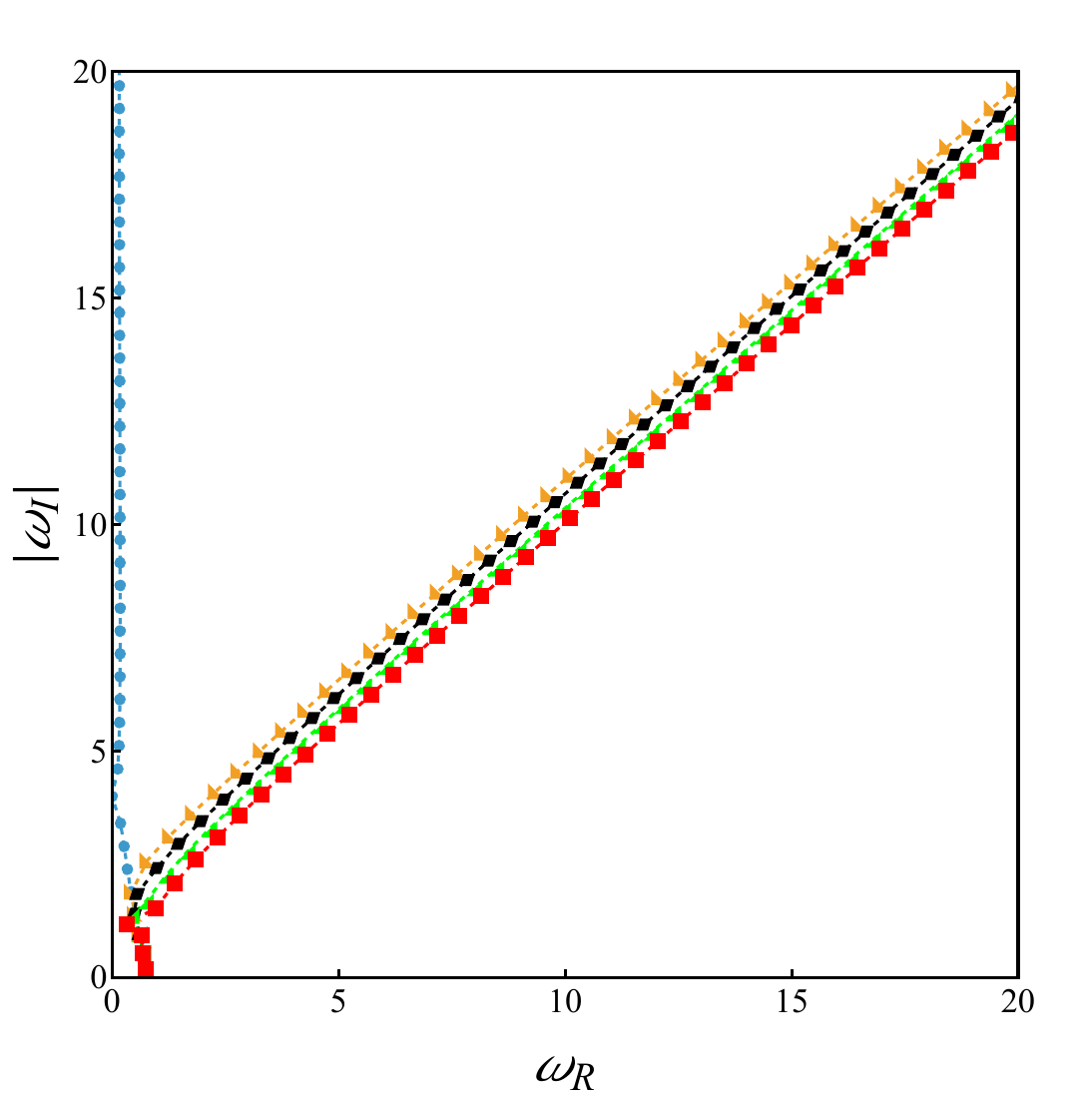}
    \caption{The first few low-lying gravitational QNM spectra, shown in the left panel, of the modified Regge-Wheeler potential Eq.~\eqref{Veff_MRW}, with various mass discontinuities at $r_c=3$.  In the right panel, a zoomed out view shows the asymptotic QNMs for different $\delta M$ are parallel.  All quantities are expressed in units of $r_h^-$.}
    \label{Fig: RW-dm-change}
\end{figure}

In Tab.~\ref{Tab. 2.5} and Fig.~\ref{Fig: RW-dm-change}, we examine the difference in the QNM spectrum between the modified Regge-Wheeler potential Eq.~\eqref{Veff_MRW} and the unperturbed potential Eq.~\eqref{V_RW} while varying $\delta M$.  
The lowest-lying modes are very close to those of the unperturbed case.  
This behavior contrasts sharply with that of the higher overtones, which are significantly displaced from their original locations, and the deviations become more pronounced for higher overtones.  
However, the deviations of the high overtones tend to converge toward a particular asymptotic behavior that is independent of the value of $\delta M$ or even the extreme case of truncation.  
Specifically, we observe that the slope of the parallel lines joining the high overtones and the spacing between consecutive modes appear to be independent of $\delta M$.  
In light of the results presented in Eqs.~\eqref{Jodinfit}-\eqref{Jodinfit2}, the asymptotic behavior described above appears to be universal and insensitive to the detailed form of the discontinuity in the effective potential.

It is worth noting that the jump discontinuity in the Regge-Wheeler potential may carry physical relevance.  
For example, a dark matter spike is a region surrounding a black hole that contains a high density of dark matter, beginning at some minimum radius and extending to a distance on the order of a kiloparsec.  
It has been shown in~\cite{agr-dark-matter-24, agr-dark-matter-59} that the minimum radius of such a spike lies between $2r_h$ and $4r_h$.  
The discontinuity in dark matter density at this minimum radius induces a discontinuity in the effective potential~\cite{agr-dark-matter-73, agr-dark-matter-75}, similar to the model examined in this work.  
In this context, we perform a detailed comparison between the low-lying modes of a Schwarzschild black hole and those of one exhibiting a discontinuity in the effective potential.  
Table~\ref{Tab.FundModeDifference} presents the results for the case of $r_c = 3r_h^-$, aimed at investigating whether such a discontinuity can be detected in the first few modes of the QNM spectrum.  
It is observed that the difference becomes more pronounced by roughly an order of magnitude for every $10\times$ increase in $\delta M$.  
Moreover, for a fixed $\delta M$, the deviation from the Schwarzschild case in each successive overtone increases by approximately an order of magnitude.

\section{Further discussions and concluding remarks}\label{sec5}

Studies based on the continued fraction method have shown that the high overtones of the original Regge-Wheeler potential asymptotically line up parallel to the imaginary frequency axis according to Eq.~\eqref{asSch}.  
The emergence of spectral instability leads to drastic deformation in the QNM spectrum, particularly the high overtones.  
The discontinuity introduced into the effective potential can be interpreted as a small-scale (or ultraviolet) perturbation to the spacetime metric. 
Existing studies on toy models have shown that the perturbed effective potential causes high overtones to asymptotically lean toward the real frequency axis, as given by Eqs.~\eqref{asTruWKB} and~\eqref{asTruWKB2}. 

For a discontinuously perturbed Regge-Wheeler potential, the above results call for an analytic assessment similar to the seminal works of Motl and Neitzke~\cite{agr-qnm-40, agr-qnm-continued-fraction-23}, which established the properties of asymptotically high overtones in the unperturbed Regge-Wheeler potential, originally demonstrated by Nollert's numerical calculations \cite{agr-qnm-continued-fraction-12}.  
This challenge is far from straightforward.  
Motl's approach~\cite{agr-qnm-continued-fraction-23} relies on a series expansion whose order is sufficiently high to ensure a convergent wavefunction yet remains lower than the overtone index.  
It remains unclear whether this approach is viable for effective potentials with discontinuities, and, to the best of our knowledge, blindly applying it does not yield meaningful results.

Primarily from the numerical perspective, the present study explores the properties of asymptotic QNMs in effective potentials featuring discontinuities.
These discontinuities have physical relevance as they mimic the presence of a thin mass shell wrapping around the black hole.
To this end, we extend the original continued fraction method to effective potentials that exhibit discontinuities.
The modified method is shown to address not only the low-lying modes but also the high overtones with high precision. 
The latter are mathematically more challenging and not accessible by most approaches for QNMs.
Specifically, the proposed algorithm comprises two key elements.
Firstly, an expansion about the horizon may introduce redundancy in the system of equations and yield meaningless results, as explained in the main text.  
We therefore perform the expansion around the point of discontinuity instead. 
Secondly, the recurrence relations for the expansion coefficients must be appropriately modified to account for the junction conditions.
The proposed algorithm is then employed to compute the QNM spectra of the modified Regge-Wheeler potential up to the \(2000^{\text{th}}\) mode, on par with Nollert’s analysis of the unperturbed Regge-Wheeler potential.
For the low-lying modes, the numerical results agree well with those obtained by using the matrix method and the Prony method.

On the analytic side, the asymptotic behavior of the deformed QNMs agrees with existing analytic assessments, primarily those derived for the modified Pöschl-Teller effective potential~\cite{agr-qnm-instability-65} given in Eq.~\eqref{asTruWKB2}.
The present study provides further details on the properties of these asymptotic high overtones.  
More specifically, in Eqs.~\eqref{Jodinfit} and~\eqref{Jodinfit2}, it is observed that the asymptotic behavior of $\Delta \omega_R$ is proportional to $1/r_c$ which seems to be universal across all known examples of spectral instability~\cite{agr-qnm-instability-65}, as is the tendency of the QNM spectrum to tilt toward the real axis with increasing $r_c$~\cite{agr-qnm-instability-02, agr-qnm-instability-03,agr-qnm-instability-11, agr-qnm-lq-03}.  

Last but not least, we discuss the possible observational implications of our findings, focusing primarily on their potential impact on gravitational-wave detection.  
The central idea is that spectral instability may leave a detectable imprint on gravitational-wave signals, as speculated in the literature~\cite{agr-qnm-instability-13}.  
This effect can be understood in terms of asymptotic modes aligning toward the real axis, potentially leading to a significant collective contribution~\cite{agr-qnm-echoes-20, agr-qnm-instability-16, agr-qnm-instability-56, agr-qnm-instability-65, agr-qnm-instability-70, agr-qnm-instability-71, agr-qnm-instability-72, agr-qnm-instability-83}.  
Moreover, the properties of high overtones are closely related to the fluid modes excited in the context of stellar QNMs~\cite{agr-qnm-star-09, agr-qnm-star-20} and to the perturbative dynamics of supermassive black holes surrounded by matter~\cite{agr-EMRI-44}.  
Regarding observations, explicit calculations of the signal-to-noise ratio~\cite{agr-TDI-review-02, agr-TDI-Wang-18}, based on realistic data streams constructed with time-delay interferometry solutions~\cite{agr-TDI-Wang-09, agr-TDI-Wang-22, agr-TDI-50} tailored to specific LISA~\cite{agr-LISA-01}, TianQin~\cite{agr-TianQin-01}, and Taiji~\cite{agr-Taiji-01} configurations, are particularly relevant.  
Such efforts are currently under active development, and preliminary results~\cite{agr-TDI-52} indicate sizable effects on model-parameter inference.  
The extent to which these effects may present additional challenges for ongoing black hole spectroscopy programs remains an open question~\cite{agr-BH-spectroscopy-review-04}.  
In addition, the study of greybody factors and Regge poles~\cite{agr-qnm-Regge-13, agr-qnm-instability-60, agr-qnm-instability-61, agr-qnm-Regge-14}, as well as reflectionless modes~\cite{agr-qnm-instability-63, agr-qnm-echoes-50}, in the presence of a discontinuous effective potential, may turn out to be intriguing, since some of these quantities are notably more stable than QNMs.

\section*{Acknowledgements}

We extend our gratitude to Qiyuan Pan and Stefan Randow for their insightful discussions.
We gratefully acknowledge the financial support from Brazilian agencies 
Funda\c{c}\~ao de Amparo \`a Pesquisa do Estado de S\~ao Paulo (FAPESP), 
Funda\c{c}\~ao de Amparo \`a Pesquisa do Estado do Rio de Janeiro (FAPERJ), 
Conselho Nacional de Desenvolvimento Cient\'{\i}fico e Tecnol\'ogico (CNPq), 
and Coordena\c{c}\~ao de Aperfei\c{c}oamento de Pessoal de N\'ivel Superior (CAPES).
This work is supported by the National Natural Science Foundation of China (NSFC).
GRL is supported by the China Scholarship Council.
A part of this work was developed under the project Institutos Nacionais de Ci\^{e}ncias e Tecnologia - F\'isica Nuclear e Aplica\c{c}\~{o}es (INCT/FNA) Proc. No. 464898/2014-5.
This research is also supported by the Center for Scientific Computing (NCC/GridUNESP) of S\~ao Paulo State University (UNESP).

\appendix

\section{The recurrence relations in the hyperboloidal coordinate}\label{app2}

In this Appendix, we give an account of the recursive relations for the wavefunction in the hyperboloidal coordinates for the effective potentials considered in this study. 

For the original Regge-Wheeler effective potential Eq.~\eqref{V_RW}, the starting point is the master equation~\eqref{CFModPsiDef0}.
By substituting the expansion Eq.~\eqref{CFModPsiDef} into Eq.~\eqref{CFModPsiDef0}, one finds
\bqn
&&\alpha_n=(1 + n) (1+n-2i\omega) ,\nb\\
&&\beta_n=-1 - \ell (1 + \ell) - 2 n (1 + n) + s^2 + 4 i \omega + 8 i n \omega + 8 \omega^2 ,\nb\\
&&\gamma_n=(n - s - 2i\omega) (n + s - 2i\omega) ,
\eqn
which is readily shown to be identical to Eq.~\eqref{threerecu1}.

Now let us consider the effective potential that contains a discontinuity at $r=r_c$.
The latter transforms to $x_c^-$ in hyperboloidal coordinates for the region to the left of the point of discontinuity, and ${x}_c^+$ for the region to the right of the discontinuity.
As discussed in the main text, in general $x_c^-\ne {x}_c^+$.

By applying the subsitution $r_h\to r_h^\pm$ (for $r\gtrless r_c$) to the height function Eq.~\eqref{HeightCompac}, one transforms the radial coordinate into the compactificed coordinate $x=1-\sigma^\pm=1-r^\pm_h/r$.
Subsequently, one adopts the following expansion for the resulting {\it stationary} wavefunction:
\bqn
\overline{\Psi}(x) = \begin{cases}
   \ a_n^- \left(\frac{x-x_c^-}{1-k(x-1)-x_c^-}\right)^n, &  x \le x_{c}^-, \\
   \ {a}_n^+ \left(\frac{x-x_c^+}{x+kx-x_c^+}\right)^n, &  x > {x}_{c}^+ ,
\end{cases} 
\lb{solhyp}
\eqn
which is obtained by factoring out the temporal factor regarding the variable $\tau$.
Here, the constant $k$ is an arbitrary real number, which is introduced to guarantee the convergence of the expansion. 
Alternatively, one might transform the expansions Eqs.~\eqref{Veff_MRWTrctdAA} and~\eqref{Veff_MRWTrctdAA1} into the hyperboloidal coordinate, this would lead to the recurrence relations governed by Eqs.~\eqref{SixTermLHSCoefficients} and~\eqref{FiveTermRHSCoefficients} presented in Sec.~\ref{sec3}.

By substituting the above expansion into the master equation~\eqref{pt_homo_eq}, both expansions lead to the following six-term recurrence relations Eqs.~\eqref{SixTermHBtruncAA}.
For $x\le x_c^-$, we have $r_h^-=2M$, and the coefficients $\alpha_n^-, \beta_n^-$, $\gamma_n^-$, $\delta_n^-$, $\epsilon_n^-$, and $\zeta_n^-$ are given by
\bqn
&&\alpha_n^-=n (1 + n) (-1 + {x_c^-}) {x_c^-},\nb\\
&&\beta_n^-=n (-1 + {x_c^-}) \big(-3 k {x_c^-} + n (1 + k - 3 {x_c^-} + 2 k {x_c^-})\big) + 
 2 i (1 + k) {r_h^-} n \big(1 + 2 (-2 + {x_c^-}) {x_c^-}\big) \omega, \nb\\
&&\gamma_n^-=k \Big(-\big((-1 + {x_c^-}) (-1 + 2 \ell (1 + \ell) + n (3 - 15 {x_c^-}) + 2 s^2 (-1 + {x_c^-}) + 7 {x_c^-} + 6 n^2 {x_c^-})\big) \nb\\
&&~~~~~~~~- 2 i {r_h^-} \big(-1 + 2 (-2 + {x_c^-}) {x_c^-} + n (5 + 2 (-2 + {x_c^-}) {x_c^-})\big) \omega - 8 {r_h^-}^2 (-2 + {x_c^-}) (-1 + {x_c^-}) \omega^2\Big) \nb\\
&&~~~~~~~~+k^2 \Big((-1 + {x_c^-})\big (2 - \ell (1 + \ell) - 5 n + 2 n^2 + s^2 + (-2 + n - s) (-2 + n + s) {x_c^-} \big) \nb\\
&&~~~~~~~~+ 2 i {r_h^-} (-1 - 4 (-2 + {x_c^-}) {x_c^-} + n (-1 + 2 (-2 + {x_c^-}) {x_c^-})) \omega - 4 {r_h^-}^2 (-2 + {x_c^-}) (-1 + {x_c^-}) \omega^2\Big)  \nb\\
&&~~~~~~~~- (-1 + {x_c^-}) \Big(\ell + \ell^2 + n^2 (2 - 3 {x_c^-}) + (-1 + s^2) (-1 + {x_c^-}) - 4 i {r_h^-} (-1 + {x_c^-}) \omega\nb\\
&&~~~~~~~~+ 4 {r_h^-}^2 (-2 + {x_c^-}) \omega^2 + n \big(-2 + 3 {x_c^-} + 8 i {r_h^-} (-1 + {x_c^-}) \omega\big)\Big), \nb\\
&&\delta_n^-=-(-1 + {x_c^-})^2 (-1 + n - s - 2 i {r_h^-} \omega) (-1 + n + s - 2 i {r_h^-} \omega) -
  k (-1 + {x_c^-}) \Big(10 + \ell + \ell^2\nb\\ 
&&~~~~~~~~- 12 n + 3 n^2 + 2 s^2 - 16 {x_c^-} + 21 n {x_c^-} - 6 n^2 {x_c^-} - 2 s^2 {x_c^-} + 4 i {r_h^-} (-4 + n) (-1 + {x_c^-}) \omega \nb\\ 
&&~~~~~~~~+ 4 {r_h^-}^2 (1 - 2 {x_c^-}) \omega^2\Big) + k^3 \Big(-\big((3 + \ell - n) (-2 + \ell + n) (-1 + {x_c^-})\big) - 6 i {r_h^-} (-2 + n) \omega\nb\\ 
&&~~~~~~~~ + 4 {r_h^-}^2 (-1 + {x_c^-}) \omega^2\Big)+ k^2 \Big(-\big((-1 + {x_c^-}) (5 + 2 \ell (1 + \ell) + 3 (-3 + n) n + s^2 + 19 {x_c^-} \nb\\
&&~~~~~~~~ + 3 (-5 + n) n {x_c^-} - s^2 {x_c^-})\big)- 2 i {r_h^-} (n (7 + 4 (-2 + {x_c^-}) {x_c^-}) - 2 (8 + 5 (-2 + {x_c^-}) {x_c^-})) \omega \nb\\
&&~~~~~~~~ + 4 {r_h^-}^2 (-1 + {x_c^-}^2) \omega^2\Big),\nb\\
&&\epsilon_n^-= k (-3 + n) \Big((-1 + {x_c^-}) \big(-2 k^2 (-3 + n) - (-3 + 2 n) (-1 + {x_c^-}) + 3 k (1 + (-3 + n) {x_c^-})\big)\nb\\
&&~~~~~~~~- 2i(1 + k) {r_h^-} \big(k^2 - 2 (-1 + {x_c^-})^2\big) \omega\Big),\nb\\
&&\zeta_n^-= k^2 (-4 + n) (-3 + n) (1 + k - {x_c^-}) (-1 + {x_c^-})
.\lb{SixTermHBtrunc1}
\eqn
For $x>{x}_c^+$, we have $r_h^+=2(M+\delta M)$.
The corresponding coefficients, denoted by $\alpha_n^+, \beta_n^+$, $\gamma_n^+$, $\delta_n^+$, $\epsilon_n^+$, and $\zeta_n^+$, have the following forms
\bqn
&&\alpha_n^+=n (1 + n) (-1 + {x_c^+})^2,\nb\\
&&\beta_n^+=-n \Big(k \big(3 + 2 n (-2 + {x_c^+}) - 3 {x_c^+}\big) + (-3 + 5 n) (-1 + {x_c^+})\Big) (-1 + {x_c^+}) -2 i k {r_h^+} n \big(1 + 2 (-2 + {x_c^+}) {x_c^+}\big)\omega, \nb\\
&&\gamma_n^+=2 (-1 + n) (-6 + 5 n) (-1 + {x_c^+})^2 + 
 k (-1 + n) \Big(\big(21 + 8 n (-2 + {x_c^+}) - 15 {x_c^+}\big) (-1 + {x_c^+}) \nb\\
&& ~~~~~~~~+ 6 i {r_h^+} \big(1 + 2 (-2 + {x_c^+}) {x_c^+}\big) \omega\Big) + k^2 \Big(9 + n^2 (6 + (-6 + {x_c^+}) {x_c^+}) - 6 i {r_h^+} \omega \nb\\
&& ~~~~~~~~- {x_c^+} \big(13 + \ell (1 + \ell) + s^2 (-1 + {x_c^+}) + 4 {x_c^+} (i + {r_h^+} \omega)^2 - 4 {r_h^+} \omega (5 i + 2 {r_h^+} \omega)\big) \nb\\
&& ~~~~~~~~+ n \big(-15 + 18 {x_c^+} - 4 {x_c^+}^2 + 2 i {r_h^+} (3 + 2 (-4 + {x_c^+}) {x_c^+}) \omega\big)\Big), \nb\\
&&\delta_n^+=-2 (-2 + n) (-9 + 5 n) (-1 + {x_c^+})^2 - 3 k (-2 + n) \Big(\big(15 + 4 n (-2 + {x_c^+}) - 9 {x_c^+}\big) (-1 + {x_c^+}) \nb\\
&& ~~~~~~~~+ 2 i {r_h^+} \big(1 + 2 (-2 + {x_c^+}) {x_c^+}\big) \omega\Big) + k^2 \Big(3 n (-4 + {x_c^+}) (-6 + 5 {x_c^+}) - 3 n^2 \big(6 + (-6 + {x_c^+}) {x_c^+}\big) \nb\\
&& ~~~~~~~~-4 i {r_h^+} n \big(3 + 2 (-4 + {x_c^+}) {x_c^+}\big)  \omega + 24 i (3 i + {r_h^+} \omega) + {x_c^+} \big(85 + \ell + \ell^2 + s^2 (-1 + {x_c^+}) - 19 {x_c^+} \nb\\
&& ~~~~~~~~+ 4 i {r_h^+} (-17 + 5 {x_c^+})  \omega + 4 {r_h^+}^2 (-2 + {x_c^+})  \omega^2\big)\Big) + k^3 \Big(-18 + 2 n^2 (-2 + {x_c^+}) + 12 i {r_h^+}  \omega \nb\\
&& ~~~~~~~~+n \big(17 - 10 {x_c^+} + 2 i {r_h^+} (-3 + 4 {x_c^+})  \omega\big) + 
    {x_c^+} \big(13 + \ell + \ell^2 - s^2 - 4 {r_h^+}  \omega (5 i + 2 {r_h^+} \omega)\big)\Big),\nb\\
&&\epsilon_n^+= (1 + k) (-3 + n) \Big((1 + k - {x_c^+})\big ((1 + k) \big(-3 (4 + k) + (5 + k) n\big) + \big(12 - 3 k (-3 + n) - 5 n\big) {x_c^+}\big) \nb\\
&& ~~~~~~~~ + 2 i k {r_h^+} \big((1 + k)^2 - 4 (1 + k) {x_c^+} + 2 {x_c^+}^2\big) \omega\Big),\nb\\
&&\zeta_n^+=-(1 +k)^2 (-4 + n) (-3 + n) (1 +k - {x_c^+})^2 
.\lb{SixTermHBtrunc2}
\eqn
Apparently, the above relations are different from Eqs.~\eqref{SixTermLHSCoefficients} and~\eqref{FiveTermRHSCoefficients}, but they are mathematically equivalent according to the discussions in~\cite{agr-qnm-hyperboloidal-22}.

In theory, the junction condition Eq.~\eqref{WronskianZero} remains unchanged.
Since for both expansions, we have $\alpha_0^-=\beta_0^-=\alpha_0^+=\beta_0^+=0$,
the junction condition, again, can be used to find the ratios between $a_0^{\pm}$ and $a_1^{\pm}$ to develop a relation between the two expansions, reminiscent of the continued fraction method.
For the present case, we must reiterate this condition Eq.~\eqref{Wronskianxc} in the hyperboloidal coordinates at the point of discontinuity, leading to some subtlety. 
Besides $x_c^+\ne x_c^-$, particular caution must be taken due to the coordinate transformation implemented by the replacements $H\to {H}^\pm$ and $G\to {G}^\pm$ for the regions $x \gtrless {x}^\pm_c$.
As a matter of fact, the procedure effectively factors out the asymptotic wavefunctions~\cite{agr-qnm-hyperboloidal-22}
\bqn
\psi^\pm_{\rm{asp}}=(r-r_h^{\pm})^{-i\omega r_h^{\pm}}r^{2i\omega r_h^{\pm}}e^{i\omega(r-r_h^{\pm})} ,\label{aspWF}
\eqn
which must be explicitly compensated for by the junction condition.

\bibliographystyle{JHEP}
\bibliography{references_qian}

\newpage

\begin{figure}[]
  \centering
  \begin{minipage}[b]{0.6\textwidth}
    \includegraphics[width=\textwidth]{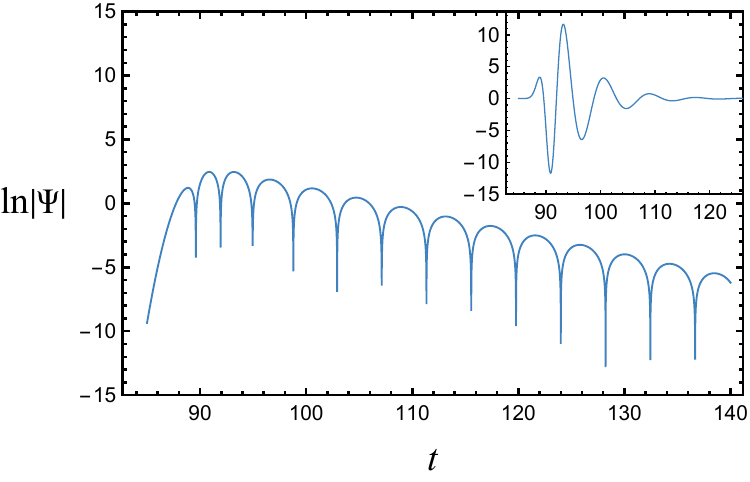}
  \end{minipage}
  \caption{Ringdown waveform $\ln|\Psi|$ as a function of $t$ for the case $\delta M = 0.005 r_h^-$ and $r_c = \frac{3}{2}~r_h^-$.  The inset shows the same waveform plotted on a linear scale.}
  \lb{Ringdown}
\end{figure}

\begin{table}[htbp] 
\caption{\label{Tab.1} Gravitational QNMs for truncated Regge-Wheeler potential Eq.~\eqref{Veff_MRWTrctd2} with $M=1/2$ and $\ell = 2$.}
\renewcommand{\arraystretch}{0.9}
\begin{tabular}{cccc}
	\hline 	\hline
	$n$ & $r_c = 3/2=1.5$& $r_c = 7/4=1.75$& $r_c = 2$\\
	\hline
1 & 0.5055960 - 0.4218217 i & 0.6444376 - 0.3957966 i & 0.7259507 - 0.3551795 i\\
2 & 0.3393478 - 1.2959620 i & 0.6331188 - 1.2233889 i & 0.8141469 - 1.1089266 i\\
3 & 0.2625940 - 2.2641040 i & 0.7435495 - 2.1389991 i & 1.03539630 - 1.9339381 i\\
4 & 0.1975060 - 3.2764305 i & 0.9312128 - 3.09982241 i & 1.3435621 - 2.7757842 i\\
5 & 0.3670901 - 4.6091741 i & 1.2374580 - 4.09344908 i & 1.7266124 - 3.6113167 i\\
6 & 0.6543376 - 5.5804341 i & 1.6178501 - 5.003107316 i & 2.1552009 - 4.4073576 i\\
7 & 0.8883882 - 6.5389791 i & 1.9899688 - 5.8689341 i & 2.5923517 - 5.1712038 i\\
8 & 1.1103650 - 7.4929249 i & 2.3556507 - 6.7202497 i & 3.02908849 - 5.9189750 i\\
9 & 1.3288173 - 8.4442428 i & 2.7189341 - 7.5648404 i & 3.4651626 - 6.6583332 i\\
10 & 1.5465540 - 9.3937729 i & 3.08167390 - 8.4055150 i & 3.9011680 - 7.3926695 i\\
20 & 3.7676977 - 18.8449132 i & 6.7448439 - 16.7331356 i & 8.2923126 - 14.6371309 i\\
30 & 6.05455880 - 28.2643246 i & 10.4637742 - 25.01050852 i & 12.7303444 - 21.8213015 i\\
40 & 8.3750159 - 37.6717165 i & 14.2117810 - 33.2697267 i & 17.1932251 - 28.9841976 i\\
50 & 10.7148574 - 47.07277796 i & 17.9766648 - 41.5194019 i & 21.6706278 - 36.1359425 i\\
60 & 13.06722778 - 56.4699434 i & 21.7524876 - 49.7631882 i & 26.1574738 - 43.2808049 i\\
70 & 15.4283405 - 65.8644753 i & 25.5359558 - 58.002978718 i & 30.6509320 - 50.4209956 i\\
80 & 17.7958923 - 75.2571108 i & 29.3250610 - 66.2398838 i & 35.1492713 - 57.5578088 i\\
90 & 20.1683803 - 84.6483170 i & 33.1184907 - 74.4746075 i & 39.6513582 - 64.6920668 i\\
100 & 22.5447711 - 94.03840824 i & 36.9153412 - 82.7076251 i & 44.1564112 - 71.8243250 i\\
200 & 46.4221651 - 187.9077352 i & 74.9834068 - 164.9891842 i & 89.2933473 - 143.08982355 i\\
300 & 70.3900585 - 281.7525459 i & 113.1309224 - 247.2327365 i & 134.4992818 - 214.3106025 i\\
400 & 94.3948533 - 375.5875939 i & 151.3108467 - 329.4610433 i & 179.7333698 - 285.5134052 i\\
500 & 118.4197917 - 469.4173670 i & 189.5084627 - 411.6810828 i & 224.9828260 - 356.7064495 i\\
1000 & 238.6757070 - 938.5321673 i & 380.6117837 - 822.7277028 i & 451.3301918 - 712.6083689 i\\
1500 & 359.02045831 - 1407.6240538 i & 571.7931063 - 1233.7381789 i & 677.7452802 - 1068.4675496 i\\
2000 & 479.4013731 - 1876.7066535 i & 763.006173758 - 1644.7339719 i & 904.1879197 - 1424.3093577 i\\
	\hline
\end{tabular}
\end{table}

\renewcommand{\arraystretch}{0.9}
\begin{table}[htbp] 
\caption{\label{Tab.2} (Continuation) Gravitational QNMs for truncated Regge-Wheeler potential Eq.~\eqref{Veff_MRWTrctd2} with $M=1/2$ and $\ell = 2$.}
\begin{tabular}{cccc}
	\hline 	\hline
	$n$ & $r_c = 3$ & $r_c = 5$ & $r_c = 10$\\
	\hline
	1 & 0.8168358 - 0.2166884 i & 0.7498252 - 0.1232622 i & 0.7072935 - 0.1152650 i\\
2 & 1.03757342 - 0.7037557 i & 0.9411521 - 0.3251548 i & 0.8568438 - 0.1687889 i\\
3 & 1.3808831 - 1.2258301 i & 1.2435854 - 0.5860281 i & 1.05510740 - 0.2681341 i\\
4 & 1.7880060 - 1.7401152 i & 1.5835126 - 0.8420711 i & 1.2719266 - 0.3589276 i\\
5 & 2.2301771 - 2.2390328 i & 1.9442554 - 1.08873699 i & 1.4965322 - 0.4455274 i\\
6 & 2.6925186 - 2.7222745 i & 2.3167763 - 1.3268562 i & 1.7258978 - 0.5287903 i\\
7 & 3.1657449 - 3.1918772 i & 2.6964107 - 1.5579378 i & 1.9583425 - 0.6093296 i\\
8 & 3.6443868 - 3.6509891 i & 3.08050628 - 1.7833422 i & 2.1928831 - 0.6876413 i\\
9 & 4.1256902 - 4.1025454 i & 3.4674586 - 2.004196025 i & 2.4289069 - 0.7641147 i\\
10 & 4.6084064 - 4.5487262 i & 3.8562697 - 2.2214116 i & 2.6660109 - 0.8390550 i\\
20 & 9.4638276 - 8.8924365 i & 7.7769428 - 4.2964049 i & 5.06116413 - 1.5405188 i\\
30 & 14.3455564 - 13.1654800 i & 11.7133824 - 6.3078515 i & 7.4682099 - 2.2038610 i\\
40 & 19.2411863 - 17.4148568 i & 15.6558192 - 8.2985959 i & 9.8777696 - 2.8532498 i\\
50 & 24.1451674 - 21.6520125 i & 19.6019270 - 10.2789739 i & 12.2884849 - 3.4957107 i\\
60 & 29.05465840 - 25.8816635 i & 23.5505196 - 12.2530829 i & 14.6999399 - 4.1340574 i\\
70 & 33.9680402 - 30.1062306 i & 27.5008984 - 14.2229811 i & 17.1119305 - 4.7696786 i\\
80 & 38.8843095 - 34.3271248 i & 31.4526191 - 16.1898527 i & 19.5243314 - 5.4033598 i\\
90 & 43.8028038 - 38.5452414 i & 35.4053827 - 18.1544433 i & 21.9370577 - 6.03558901 i\\
100 & 48.7230637 - 42.7611840 i & 39.3589790 - 20.1172529 i & 24.3500484 - 6.6666900 i\\
200 & 97.9773020 - 84.8585292 i & 78.9195427 - 39.6946902 i & 48.4880431 - 12.9458695 i\\
300 & 147.2728915 - 126.9071958 i & 118.4999521 - 59.2325453 i & 72.6327717 - 19.2003474 i\\
400 & 196.5853703 - 168.9362738 i & 158.08849301 - 78.7545174 i & 96.7803166 - 25.4449634 i\\
500 & 245.9070647 - 210.9547127 i & 197.6814685 - 98.2678742 i & 120.9294038 - 31.6842413 i\\
1000 & 492.5754856 - 420.9778626 i & 395.6751293 - 195.7788202 i & 241.6848416 - 62.8461055 i\\
1500 & 739.2844060 - 630.9543849 i & 593.6881696 - 293.2521041 i & 362.4469784 - 93.9847306 i\\
2000 & 986.009771507 - 840.9119518 i & 791.7090469 - 390.7100952 i & 483.2118018 - 125.1139383 i\\
	\hline
\end{tabular}
\end{table}

\renewcommand{\arraystretch}{0.9}
\begin{table}[htbp] 
\caption{\label{Tab.3} (Continuation) Gravitational QNMs for truncated Regge-Wheeler potential Eq.~\eqref{Veff_MRWTrctd2} with $M=1/2$ and $\ell = 2$.}
\begin{tabular}{cccc}
	\hline 	\hline
	$n$ & $r_c = 15$ & $r_c = 25$ & $r_c = 35$\\
	\hline
		1 & 0.3380131 - 0.08913340 i & 0.2108654 - 0.05865160 i & 0.1530649 - 0.04344197 i\\
2 & 0.5208478 - 0.1028967 i & 0.3309380 - 0.07085257 i & 0.2418243 - 0.05309889 i\\
3 & 0.6813897 - 0.1044864 i & 0.4441582 - 0.07835949 i & 0.3263675 - 0.05946229 i\\
4 & 0.8070321 - 0.1217418 i & 0.5530046 - 0.08295471 i & 0.4087395 - 0.06407702 i\\
5 & 0.9469401 - 0.1730117 i & 0.6565115 - 0.08505821 i & 0.4894845 - 0.06749777 i\\
6 & 1.1033229 - 0.2219358 i & 0.7506358 - 0.08901834 i & 0.5686099 - 0.06993562 i\\
7 & 1.2647737 - 0.2677548 i & 0.8418092 - 0.1048475 i & 0.6455393 - 0.07158111 i\\
8 & 1.4290773 - 0.3116288 i & 0.9404008 - 0.1260066 i & 0.7190257 - 0.07367017 i\\
9 & 1.5952816 - 0.3539907 i & 1.04337424 - 0.1463207 i & 0.7897492 - 0.07959246 i\\
1 & 0.3380131 - 0.08913340 i & 0.2108654 - 0.05865160 i & 0.1530649 - 0.04344197 i\\
2 & 0.5208478 - 0.1028967 i & 0.3309380 - 0.07085257 i & 0.2418243 - 0.05309889 i\\
3 & 0.6813897 - 0.1044864 i & 0.4441582 - 0.07835949 i & 0.3263675 - 0.05946229 i\\
4 & 0.8070321 - 0.1217418 i & 0.5530046 - 0.08295471 i & 0.4087395 - 0.06407702 i\\
5 & 0.9469401 - 0.1730117 i & 0.6565115 - 0.08505821 i & 0.4894845 - 0.06749777 i\\
6 & 1.1033229 - 0.2219358 i & 0.7506358 - 0.08901834 i & 0.5686099 - 0.06993562 i\\
7 & 1.2647737 - 0.2677548 i & 0.8418092 - 0.1048475 i & 0.6455393 - 0.07158111 i\\
8 & 1.4290773 - 0.3116288 i & 0.9404008 - 0.1260066 i & 0.7190257 - 0.07367017 i\\
9 & 1.5952816 - 0.3539907 i & 1.04337424 - 0.1463207 i & 0.7897492 - 0.07959246 i\\
200 & 34.5112448 - 6.4620762 i & 22.03158046 - 2.6530774 i & 16.2119818 - 1.4651847 i\\
300 & 51.7689917 - 9.5608337 i & 33.04144426 - 3.8966682 i & 24.3111114 - 2.1376668 i\\
400 & 69.02825176 - 12.6523335 i & 44.05198894 - 5.1354414 i & 32.4106422 - 2.8064685 i\\
500 & 86.2883449 - 15.7399291 i & 55.06291623 - 6.3716459 i & 40.5104025 - 3.4733265 i\\
1000 & 172.5942128 - 31.1527872 i & 110.1200650 - 12.5362566 i & 81.01073847 - 6.7952887 i\\
1500 & 258.9036837 - 46.5488077 i & 165.1788933 - 18.6899262 i & 121.5121106 - 10.1090759 i\\
2000 & 345.2145887 - 61.9380281 i & 220.2383861 - 24.8391959 i & 162.01389302 - 13.4195887 i\\
	\hline
\end{tabular}
\end{table}

\renewcommand{\arraystretch}{0.9}
\begin{table}[htbp] 
\caption{\label{Tab.4} (Continuation) Gravitational QNMs for truncated Regge-Wheeler potential Eq.~\eqref{Veff_MRWTrctd2} with $M=1/2$ and $\ell = 2$.}
\begin{tabular}{ccc}
	\hline 	\hline
	$n$ & $r_c = 50$ & $r_c = 100$ \\
	\hline
1 & 0.10842489 - 0.03121472 i & 0.05495620 - 0.01607373 i\\
2 & 0.1720927 - 0.03842984 i & 0.08766966 - 0.01993071 i\\
3 & 0.2330871 - 0.04330685 i & 0.11917987 - 0.02258258 i\\
4 & 0.2928490 - 0.04698026 i & 0.1501930 - 0.02461756 i\\
5 & 0.3518196 - 0.04989387 i & 0.1809263 - 0.02626930 i\\
6 & 0.4101777 - 0.05226828 i & 0.2114740 - 0.02765779 i\\
7 & 0.4679914 - 0.05422385 i & 0.2418847 - 0.02885351 i\\
8 & 0.5252553 - 0.05582830 i & 0.2721868 - 0.02990150 i\\
9 & 0.5818819 - 0.05713070 i & 0.3023973 - 0.03083227 i\\
10 & 0.6376665 - 0.05823165 i & 0.3325275 - 0.03166743 i\\
20 & 1.1893377 - 0.1037706 i & 0.6297681 - 0.03701650 i\\
30 & 1.7632920 - 0.1515146 i & 0.9196874 - 0.04723370 i\\
40 & 2.3411175 - 0.1947535 i & 1.2154687 - 0.06156533 i\\
50 & 2.9203703 - 0.2355610 i & 1.5134217 - 0.07457023 i\\
60 & 3.5002914 - 0.2748528 i & 1.8121925 - 0.08674895 i\\
70 & 4.08056365 - 0.3131253 i & 2.1113933 - 0.09836861 i\\
80 & 4.6610378 - 0.3506756 i & 2.4108512 - 0.1095834 i\\
90 & 5.2416365 - 0.3876907 i & 2.7104744 - 0.1204920 i\\
100 & 5.8223143 - 0.4242952 i & 3.01020941 - 0.13116147 i\\
200 & 11.6304774 - 0.7795005 i & 6.009747243 - 0.2312180 i\\
300 & 17.4393415 - 1.1270342 i & 9.01021454 - 0.3264523 i\\
400 & 23.2484403 - 1.4717775 i & 12.01083712 - 0.4199799 i\\
500 & 29.05767298 - 1.8150688 i & 15.01151908 - 0.5126525 i\\
1000 & 58.1047523 - 3.5224228 i & 30.01528237 - 0.9708619 i\\
1500 & 87.1524627 - 5.2237903 i & 45.01929357 - 1.4257814 i\\
2000 & 116.2004259 - 6.9227737 i & 60.02340902 - 1.8794172 i\\
	\hline
\end{tabular}
\end{table}

\begin{figure}[]
  \centering
  \begin{minipage}[b]{0.8\textwidth}
    \includegraphics[width=\textwidth]{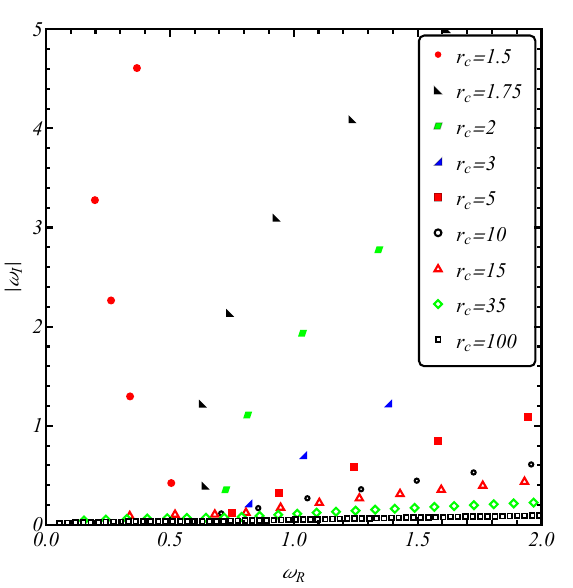}
  \end{minipage}
  \caption{\lb{RW_trunc_QNM_zoom} 
  The first few low-lying axial gravitational QNMs of the truncated effective potential Eq.~\eqref{Veff_MRWTrctd2} for different values of $r_c$, in units of $r_h^{-1}$.}
\end{figure}

\begin{figure}
  \begin{minipage}[b]{0.8\textwidth}
    \includegraphics[width=\textwidth]{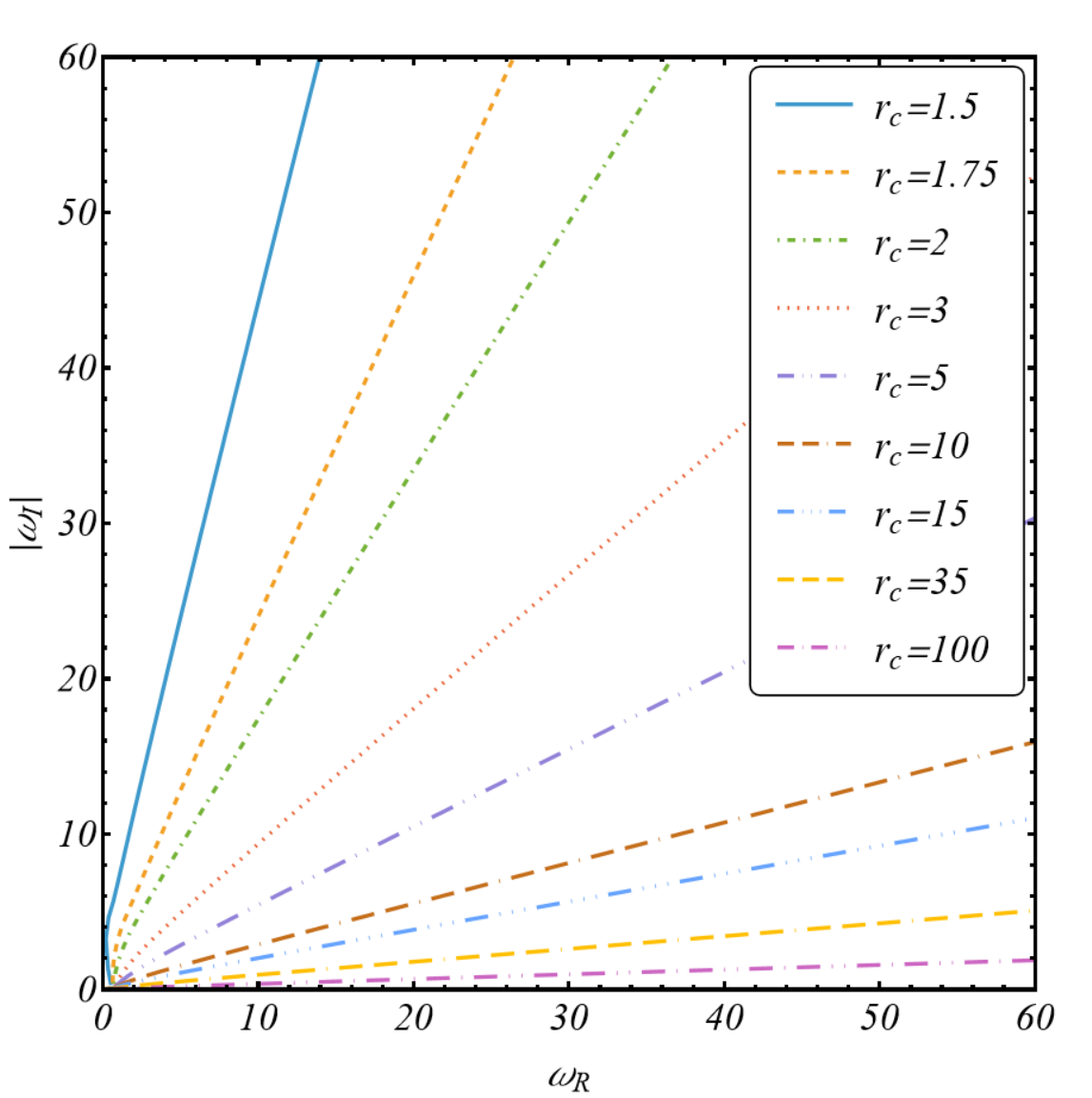}
  \end{minipage}
\caption{\lb{RW_trunc_QNM} The same as Fig.~\ref{RW_trunc_QNM_zoom} but focusing on the high overtones by showing a larger region of the complex frequency plane.  The slope decreases as $r_c$ increases.}
\end{figure}

\begin{table}[htbp] 
\caption{\label{Tab. 2.1} Gravitational QNMs for Regge-Wheeler potential with a minor step Eq.~\eqref{Veff_MRW} with $M=1/2$, $\delta M=0.005$ and $\ell = 2$.}
\renewcommand{\arraystretch}{0.9}
\begin{tabular}{cccc}
	\hline 	\hline
	$n$ & $r_c = 3/2=1.5$& $r_c = 7/4=1.75$& $r_c = 2$\\
	\hline
		1 & 0.7443680 - 0.1754761 i & 0.7449842 - 0.1768751 i & 0.7454524 - 0.1777602 i\\
2 & 0.6941990 - 0.5432699 i & 0.6932467 - 0.5435928 i & 0.6923259 - 0.5433697 i\\
3 & 0.6050793 - 0.9508322 i & 0.6032079 - 0.9525066 i & 0.6030001 - 0.9556444 i\\
4 & 0.5103523 - 1.4023587 i & 0.5093247 - 1.3999753 i & 0.5074176 - 1.3886860 i\\
5 & 0.4245862 - 1.8844595 i & 0.4170715 - 1.8922718 i & 0.4075005 - 1.9666110 i\\
6 & 0.3533941 - 2.3788182 i & 0.3602759 - 2.3599672 i & 0.3814876 - 2.1861104 i\\
7 & 0.2824926 - 2.8828259 i & 0.2335021 - 2.9317408 i & 0.7514376 - 3.01047687 i\\
8 & 0.2086102 - 3.3904749 i & 0.2887961 - 3.2929555 i & 1.1606156 - 3.9091010 i\\
9 & 0.0305779 - 3.94834 i & 0.6255822 - 4.3713450 i & 1.6280138 - 4.7135136 i\\
10 & 0.1621723 - 4.5884456 i & 1.06069129 - 5.2522053 i & 2.08040608 - 5.4739389 i\\
20 & 1.2106963 - 10.4701299 i & 4.002715513 - 11.9839350 i & 6.4747064 - 12.7767371 i\\
30 & 3.2152363 - 18.9777271 i & 7.6946119 - 20.2872024 i & 10.9036933 - 19.9746078 i\\
40 & 5.5025887 - 28.3998644 i & 11.4280946 - 28.5576298 i & 15.3614070 - 27.1438551 i\\
50 & 7.8232517 - 37.8087251 i & 15.1842143 - 36.8135433 i & 19.8355803 - 34.2992590 i\\
60 & 10.1632015 - 47.2106924 i & 18.9542156 - 45.06129393 i & 24.3202349 - 41.4464966 i\\
70 & 12.5156370 - 56.6084696 i & 22.7335477 - 53.3038235 i & 28.8121139 - 48.5883515 i\\
80 & 14.8767925 - 66.003441448 i & 26.5195673 - 61.5427316 i & 33.3092630 - 55.7263945 i\\
90 & 17.2443742 - 75.3964082 i & 30.3106089 - 69.7789829 i & 37.8104217 - 62.8615979 i\\
100 & 19.6168842 - 84.7878725 i & 34.10555753 - 78.01320347 i & 42.3147312 - 69.9946051 i\\
200 & 43.4773333 - 178.6629957 i & 72.1651837 - 160.2999604 i & 87.4483213 - 141.2633883 i\\
300 & 67.4397856 - 272.5095984 i & 110.3099429 - 242.5451628 i & 132.6531468 - 212.4852273 i\\
400 & 91.4419048 - 366.3455149 i & 148.4885036 - 324.7742778 i & 177.8866833 - 283.6885522 i\\
500 & 115.4652537 - 460.1758001 i & 186.6853070 - 406.9947964 i & 223.1358099 - 354.8819071 i\\
1000 & 235.7180309 - 929.2916029 i & 377.7870186 - 818.04236051 i & 449.4825213 - 710.7844406 i\\
1500 & 356.06175052 - 1398.3838166 i & 568.9678106 - 1229.05314653 i & 675.8973934 - 1066.6438236 i\\
2000 & 476.4421527 - 1867.4665785 i & 760.1806141 - 1640.04909357 i & 902.3399252 - 1422.4857322 i\\
	\hline
\end{tabular}
\end{table}

\renewcommand{\arraystretch}{0.9}
\begin{table}[htbp] 
\caption{\label{Tab. 2.2} (Continuation) Gravitational QNMs for Regge-Wheeler potential with a minor step Eq.~\eqref{Veff_MRW} with $M=1/2$, $\delta M=0.005$ and $\ell = 2$.}
\begin{tabular}{cccc}
	\hline 	\hline
	$n$ & $r_c = 3$ & $r_c = 5$ & $r_c = 10$\\
	\hline
1 & 0.7472389 - 0.1788025 i & 0.7473412 - 0.1774461 i & 0.7468567 - 0.1780639 i\\
2 & 0.6837598 - 0.5456597 i & 0.7371170 - 0.5443838 i & 0.4944461 - 0.3865552 i\\
3 & 0.6618900 - 0.9339999 i & 0.4795737 - 0.6434485 i & 0.7423563 - 0.3948773 i\\
4 & 0.3157503 - 1.1819340 i & 0.0397601 - 0.838888 i & 0.2148164 - 0.4239794 i\\
5 & 0.9611385 - 1.5295124 i & 1.004117746 - 0.8804742 i & 0.9580214 - 0.4874909 i\\
6 & 1.3860365 - 2.08157527 i & 1.3576044 - 1.1612163 i & 1.1808916 - 0.5872479 i\\
7 & 1.8438522 - 2.6015827 i & 1.7287572 - 1.4207302 i & 1.4102589 - 0.6793445 i\\
8 & 2.3198789 - 3.09599065 i & 2.1091008 - 1.6662229 i & 1.6431608 - 0.7660320 i\\
9 & 2.8034940 - 3.5706089 i & 2.4947326 - 1.9017699 i & 1.8783201 - 0.8488421 i\\
10 & 3.2892647 - 4.03179510 i & 2.8834763 - 2.1299373 i & 2.1150056 - 0.9287282 i\\
20 & 8.1520011 - 8.4169582 i & 6.8096377 - 4.2412117 i & 4.5115612 - 1.6501593 i\\
30 & 13.03248709 - 12.6993857 i & 10.7472647 - 6.2603704 i & 6.9199585 - 2.3185921 i\\
40 & 17.9270081 - 16.9531220 i & 14.6899011 - 8.2545038 i & 9.3301400 - 2.9701307 i\\
50 & 22.8302031 - 21.1928066 i & 18.6360104 - 10.2368088 i & 11.7411707 - 3.6137711 i\\
60 & 27.7391277 - 25.4241087 i & 22.5845594 - 12.2121651 i & 14.1528055 - 4.2528653 i\\
70 & 32.6520868 - 29.6498379 i & 26.5348871 - 14.1829369 i & 16.5649085 - 4.8890040 i\\
80 & 37.5680301 - 33.8715939 i & 30.4865594 - 16.1504547 i & 18.9773850 - 5.5230652 i\\
90 & 42.4862662 - 38.09037480 i & 34.4392798 - 18.1155424 i & 21.3901649 - 6.1555855 i\\
100 & 47.4063167 - 42.3068450 i & 38.3928382 - 20.07874635 i & 23.8031952 - 6.7869166 i\\
200 & 96.6595887 - 84.4065164 i & 77.9531999 - 39.6579180 i & 47.9413355 - 13.06710348 i\\
300 & 145.9548505 - 126.4559382 i & 117.5335330 - 59.1963344 i & 72.08610210 - 19.3219063 i\\
400 & 195.2671652 - 168.4853891 i & 157.1220347 - 78.7185833 i & 96.2336644 - 25.5666821 i\\
500 & 244.5887614 - 210.5040500 i & 196.7149865 - 98.2321047 i & 120.3827618 - 31.8060550 i\\
1000 & 491.2569867 - 420.5276392 i & 394.7086000 - 195.7433759 i & 241.1382192 - 62.9681065 i\\
1500 & 737.9658425 - 630.5043061 i & 592.7216247 - 293.2167667 i & 361.9003624 - 94.1067931 i\\
2000 & 984.6911758 - 840.4619450 i & 790.7424943 - 390.6748110 i & 482.6651891 - 125.2360312 i\\
	\hline
\end{tabular}
\end{table}

\renewcommand{\arraystretch}{0.9}
\begin{table}[htbp] 
\caption{\label{Tab. 2.3} (Continuation) Gravitational QNMs for Regge-Wheeler potential with a minor step Eq.~\eqref{Veff_MRW} with $M=1/2$, $\delta M=0.005$ and $\ell = 2$.}
\begin{tabular}{cccc}
	\hline 	\hline
	$n$ & $r_c = 15$ & $r_c = 25$ & $r_c = 35$\\
	\hline
1 & 0.7477917 - 0.1789394 i & 0.7460984 - 0.1711300 i & 0.1474100 - 0.1564691 i\\
2 & 0.523631 - 0.295479 i & 0.2051630 - 0.2052864 i & 0.7208131 - 0.1569516 i\\
3 & 0.709734 - 0.297005 i & 0.6645626 - 0.2059347 i & 0.05391037 - 0.1575107 i\\
4 & 0.33675 - 0.29989 i & 0.3261591 - 0.2074499 i & 0.2367757 - 0.1593071 i\\
5 & 0.135963 - 0.314374 i & 0.07746687 - 0.2088737 i & 0.7753788 - 0.1610527 i\\
6 & 0.8887638 - 0.3453348 i & 0.4414664 - 0.2096097 i & 0.3225392 - 0.1623440 i\\
7 & 1.05153685 - 0.3977649 i & 0.5533155 - 0.2101458 i & 0.4059995 - 0.1647692 i\\
8 & 1.2159900 - 0.4467283 i & 0.8077560 - 0.2182277 i & 0.6475805 - 0.1650988 i\\
9 & 1.3824400 - 0.4927947 i & 0.9141800 - 0.2471961 i & 0.4878290 - 0.1663892 i\\
10 & 1.5502958 - 0.5366828 i & 1.01988183 - 0.2696512 i & 0.5683703 - 0.1668917 i\\
20 & 3.2578610 - 0.9196959 i & 2.1010070 - 0.4451372 i & 1.5564034 - 0.2790825 i\\
30 & 4.9790410 - 1.2634528 i & 3.1968306 - 0.5937037 i & 2.3604545 - 0.3653121 i\\
40 & 6.7024671 - 1.5937374 i & 4.2957439 - 0.7328966 i & 3.1679424 - 0.4443677 i\\
50 & 8.4265582 - 1.9176335 i & 5.3956418 - 0.8674148 i & 3.9766409 - 0.5197817 i\\
60 & 10.1509771 - 2.2378973 i & 6.4959290 - 0.9992233 i & 4.7858584 - 0.5930331 i\\
70 & 11.8756167 - 2.5558370 i & 7.5964017 - 1.1293001 i & 5.5953264 - 0.6648803 i\\
80 & 13.6004263 - 2.8721655 i & 8.6969801 - 1.2581882 i & 6.4049279 - 0.7357570 i\\
90 & 15.3253745 - 3.1873120 i & 9.7976283 - 1.3862152 i & 7.2146073 - 0.8059299 i\\
100 & 17.05043886 - 3.5015543 i & 10.8983279 - 1.5135913 i & 8.02433631 - 0.8755726 i\\
200 & 34.3048852 - 6.6191765 i & 21.9068588 - 2.7699956 i & 16.1227064 - 1.5580791 i\\
300 & 51.5627057 - 9.7180615 i & 32.9167835 - 4.01366719 i & 24.2218880 - 2.2306215 i\\
400 & 68.8220010 - 12.8096240 i & 43.9273569 - 5.2524798 i & 32.3214430 - 2.8994522 i\\
500 & 86.08211482 - 15.8972568 i & 54.9383009 - 6.4887075 i & 40.4212172 - 3.5663273 i\\
1000 & 172.3880230 - 31.3101879 i & 109.9954820 - 12.6533637 i & 80.9215800 - 6.8883229 i\\
1500 & 258.6975071 - 46.7062322 i & 165.05432074 - 18.8070481 i & 121.4229607 - 10.2021210 i\\
2000 & 345.008418557 - 62.09546440 i & 220.1138186 - 24.9563251 i & 161.9247473 - 13.5126390 i\\
	\hline
\end{tabular}
\end{table}

\renewcommand{\arraystretch}{0.9}
\begin{table}[htbp] 
\caption{\label{Tab. 2.4} (Continuation) Gravitational QNMs for Regge-Wheeler potential with a minor step Eq.~\eqref{Veff_MRW} with $M=1/2$, $\delta M=0.005$ and $\ell = 2$.}
\begin{tabular}{ccc}
	\hline 	\hline
	$n$ & $r_c = 50$ & $r_c = 100$ \\
	\hline
1 & 0.03692594 - 0.11588026 i & 0.11688742 - 0.06647446 i\\
2 & 0.1035765 - 0.1159266 i & 0.1482814 - 0.06795937 i\\
3 & 0.1676959 - 0.1184927 i & 0.1793026 - 0.06926280 i\\
4 & 0.2295525 - 0.1212369 i & 0.2100713 - 0.07040615 i\\
5 & 0.2900099 - 0.1236355 i & 0.2406561 - 0.07141521 i\\
6 & 0.3495296 - 0.1256379 i & 0.2710986 - 0.07231214 i\\
7 & 0.4083475 - 0.1272680 i & 0.3014254 - 0.07311455 i\\
8 & 0.7427118 - 0.1276294 i & 0.3316536 - 0.07383609 i\\
9 & 0.6922331 - 0.1281734 i & 0.3617950 - 0.07448726 i\\
10 & 0.4665848 - 0.1285394 i & 0.3918571 - 0.07507600 i\\
20 & 1.1257121 - 0.1718612 i & 0.6874786 - 0.07834924 i\\
30 & 1.6998927 - 0.2212694 i & 0.9764008 - 0.09149700 i\\
40 & 2.2779149 - 0.2651190 i & 1.2731237 - 0.1059280 i\\
50 & 2.8573095 - 0.3062396 i & 1.5714235 - 0.1188720 i\\
60 & 3.4373347 - 0.3457191 i & 1.8703849 - 0.1309907 i\\
70 & 4.01768737 - 0.3841130 i & 2.1697065 - 0.1425592 i\\
80 & 4.5982236 - 0.4217444 i & 2.4692471 - 0.1537314 i\\
90 & 5.1788688 - 0.4588168 i & 2.7689301 - 0.1646044 i\\
100 & 5.7595819 - 0.4954652 i & 3.06870983 - 0.1752440 i\\
200 & 11.5679008 - 0.8508376 i & 6.06840951 - 0.2751559 i\\
300 & 17.3768079 - 1.1984166 i & 9.06891423 - 0.3703434 i\\
400 & 23.1859267 - 1.5431812 i & 12.06955262 - 0.4638494 i\\
500 & 28.9951708 - 1.8864849 i & 15.07024330 - 0.5565099 i\\
1000 & 58.04227163 - 3.5938631 i & 30.07402271 - 1.01469690 i\\
1500 & 87.08998892 - 5.2952383 i & 45.07803894 - 1.4696095 i\\
2000 & 116.1379554 - 6.9942255 i & 60.08215683 - 1.9232421 i\\
	\hline
\end{tabular}
\end{table}

\begin{figure}[]
  \centering
  \begin{minipage}[b]{0.6\textwidth}
    \includegraphics[width=\textwidth]{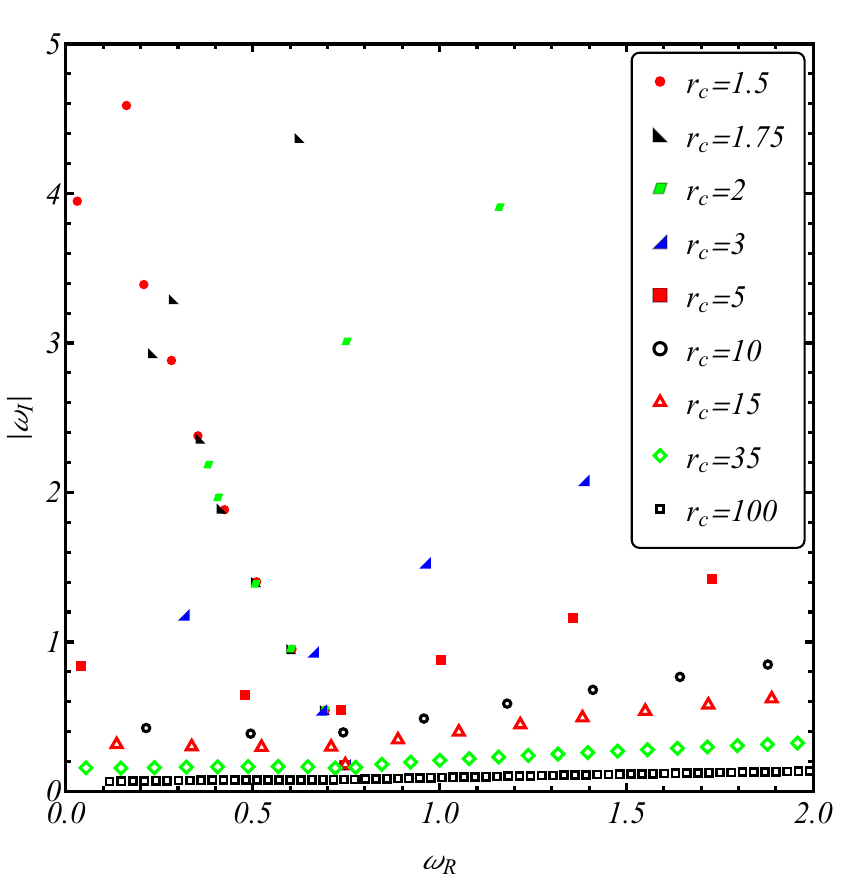}
  \end{minipage}
  \caption{ The first few low-lying gravitational QNM spectra of the modified Regge-Wheeler potential Eq.~\eqref{Veff_MRW} for different values of $r_c$ in units of $r_h^-$.}
  \lb{RW_2Sided_QNM_zoom}
\end{figure}

\begin{figure}[]
  \centering
  \begin{minipage}[b]{0.6\textwidth}
    \includegraphics[width=\textwidth]{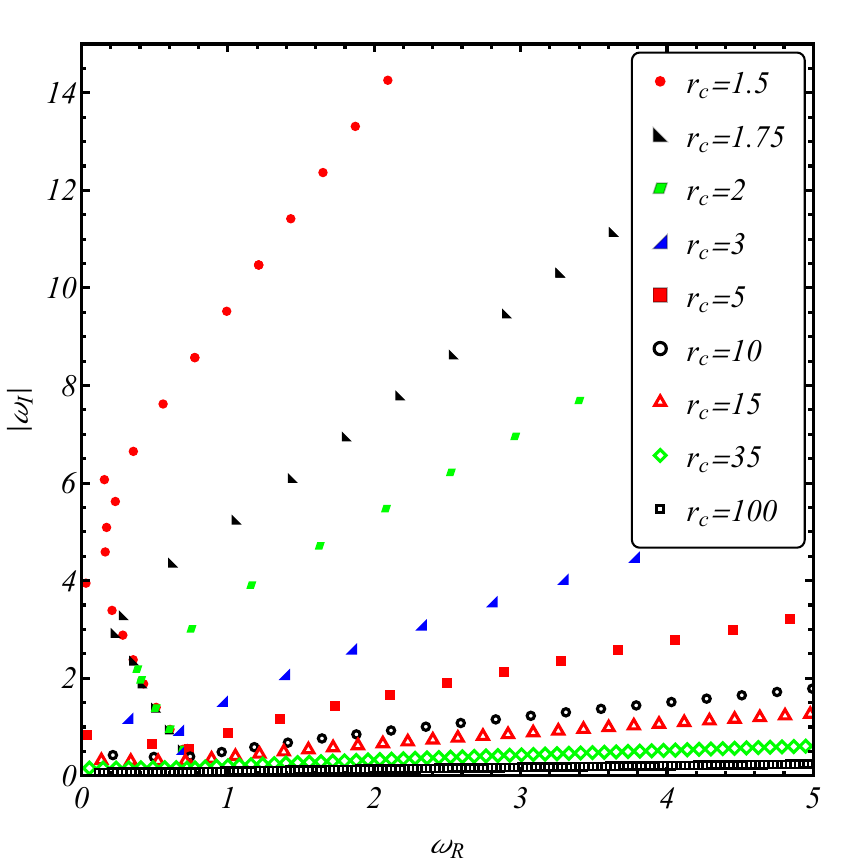}
  \end{minipage}
  \caption{ The same as Fig.~\ref{RW_2Sided_QNM_zoom} but focusing on the high overtones by showing a larger region of the complex frequency plane.}
  \lb{RW_2Sided_QNM}
\end{figure}

\begin{table}[htbp] 
\caption{\label{Tab. 2.5} Gravitational QNMs for Regge-Wheeler potential with a minor step Eq.~\eqref{Veff_MRW} with $M=1/2$, $r_{c}=3$, and $\ell = 2$.}
\resizebox{\textwidth}{!}{\begin{tabular}{ccccc}
	\hline 	\hline
	$n$ & $\delta M=0.000005$ & $\delta M=0.00005$ & $\delta M=0.0005$  & $\delta M=0.005$ \\
	\hline
1 & 0.7473433 - 0.1779255 i & 0.7473423 - 0.1779334 i & 0.7473323 - 0.1780119 i  & 0.7472389 - 0.1788025 i\\
2 & 0.6934124 - 0.5478281 i & 0.6933262 - 0.5478129 i & 0.6924633 - 0.5476570 i  & 0.6837598 - 0.5456597 i\\
3 & 0.6021832 - 0.9565467 i & 0.6028686 - 0.9564787 i & 0.6096182 - 0.9555462 i  & 0.6618900 - 0.9339999 i\\
4 & 0.5018405 - 1.4101202 i & 0.4915274 - 1.4078262 i & 0.4279066 - 1.3609576 i  & 0.3157503 - 1.1819340 i\\
5 & 0.4371802 - 1.8989328 i & 0.5529713 - 1.8469253 i & 0.7455202 - 1.6815371 i  & 0.9611385 - 1.5295124 i\\
6 & 0.7808154 - 2.5906612 i & 0.9865021 - 2.4161437 i & 1.1882671 - 2.2464370 i  & 1.3860365 - 2.08157527 i\\
7 & 1.2775014 - 3.1359216 i & 1.4683426 - 2.9530778 i & 1.6572492 - 2.7752095 i  & 1.8438522 - 2.6015827 i\\
8 & 1.7862926 - 3.6386280 i & 1.9643619 - 3.4543837 i & 2.1424498 - 3.2736077 i  & 2.3198789 - 3.09599065 i\\
9 & 2.2906470 - 4.1124063 i & 2.4613151 - 3.9298361 i & 2.6324255 - 3.7492265 i  & 2.8034940 - 3.5706089 i\\
10 & 2.7885771 - 4.5712588 i & 2.9552184 - 4.3902531 i & 3.1222146 - 4.2103951 i  & 3.2892647 - 4.03179510 i\\
20 & 7.6783645 - 8.9528026 i & 7.8363003 - 8.7740526 i & 7.9942079 - 8.5954104 i  & 8.1520011 - 8.4169582 i\\
30 & 12.5637075 - 13.2368961 i & 12.7200184 - 13.05765392 i & 12.8763043 - 12.8784567 i  & 13.03248709 - 12.6993857 i\\
40 & 17.4603242 - 17.4917451 i & 17.6159296 - 17.3121495 i & 17.7715161 - 17.1325813 i  & 17.9270081 - 16.9531220 i\\
50 & 22.3646880 - 21.7321579 i & 22.5198998 - 21.5523268 i & 22.6750963 - 21.3725158 i  & 22.8302031 - 21.1928066 i\\
60 & 27.2743572 - 25.9639646 i & 27.4293184 - 25.7839689 i & 27.5842664 - 25.6039898 i  & 27.7391277 - 25.4241087 i\\
70 & 32.1878316 - 30.1900616 i & 32.3426195 - 30.009945389 i & 32.4973955 - 29.8298436 i  & 32.6520868 - 29.6498379 i\\
80 & 37.1041525 - 34.4120968 i & 37.2588135 - 34.2318889 i & 37.4134635 - 34.05169405 i  & 37.5680301 - 33.8715939 i\\
90 & 42.02267692 - 38.6310964 i & 42.1772411 - 38.4508165 i & 42.3317948 - 38.2705487 i  & 42.4862662 - 38.09037480 i\\
100 & 46.9429548 - 42.8477424 i & 47.09744262 - 42.6674045 i & 47.2519205 - 42.4870781 i  & 47.4063167 - 42.3068450 i\\
200 & 96.1972123 - 84.9482096 i & 96.3513697 - 84.7676083 i & 96.5055187 - 84.5870166 i  & 96.6595887 - 84.4065164 i\\
300 & 145.4927882 - 126.9978961 i & 145.6468404 - 126.8172069 i & 145.8008846 - 126.6365270 i  & 145.9548505 - 126.4559382 i\\
400 & 194.8052571 - 169.02747859 i & 194.9592578 - 168.8467456 i & 195.1132505 - 168.6660218 i  & 195.2671652 - 168.4853891 i\\
500 & 244.12694468 - 211.04621808 i & 244.2809148 - 210.8654589 i & 244.4348770 - 210.6847090 i  & 244.5887614 - 210.5040500 i\\
1000 & 490.7953505 - 421.06996305 i & 490.9492603 - 420.8891520 i & 491.1031623 - 420.7083501 i  & 491.2569867 - 420.5276392 i\\
1500 & 737.5042656 - 631.04668135 i & 737.6581556 - 630.8658532 i & 737.8120378 - 630.6850342 i  & 737.9658425 - 630.5043061 i\\
2000 & 984.2297088 - 841.004485506 i & 984.3835086 - 840.8235091 i & 984.5373810 - 840.6426816 i  & 984.6911758 - 840.4619450 i\\
	\hline
\end{tabular}}
\end{table}

\begin{table}[htbp] 
\caption{\label{Tab.FundModeDifference} Difference $|\omega_{\text{Schw}}-\omega_{\text{disc}}|$ in the low-damping modes between the Schwarzschild case and the modified Regge-Wheeler potential Eq.~\eqref{Veff_MRW} with $r_c=3$, as a function of increasing $\delta M$.  
In contrast to the higher overtones, the differences in the first three modes scale linearly with $\delta M$.}
\begin{tabular}{ccccccc}
	\hline 	\hline
    $\delta M$&\multicolumn{1}{c}{$n=0$}&\multicolumn{1}{c}{$n=1$}&\multicolumn{1}{c}{$n=2$}&\multicolumn{1}{c}{$n=3$}& $\cdots$ &\multicolumn{1}{c}{$n=500$}\\
    \hline
0.000005 & 0.000000879  & 0.00000972 & 0.0000766&0.0011827& $\cdots$ &247.565\\
0.00005 & 0.00000879 & 0.0000972  & 0.000765&0.0117452& $\cdots$ &247.745\\
0.0005 & 0.0000880  & 0.000974 &0.00758&0.0898601& $\cdots$ &247.926\\
0.005 & 0.000884 & 0.00990  & 0.0634&0.295323& $\cdots$ &248.106\\
	\hline
\end{tabular}
\end{table}

\end{document}